\documentclass[compsoc,conference,a4paper,10pt,times]{IEEEtran}
\IEEEoverridecommandlockouts

\usepackage{cite}
\usepackage{amsmath,amssymb,amsfonts}
\usepackage{algorithmic}
\usepackage{graphicx}
\usepackage{textcomp}
\usepackage{bmpsize}
\usepackage{xcolor}
\usepackage{lipsum}
\usepackage[colorlinks=true,urlcolor=black]{hyperref}

\usepackage{framed}
\usepackage{multirow}
\usepackage{booktabs}
\usepackage{ifthen}
\usepackage{color}
\usepackage{url}
\usepackage{amsmath}
\usepackage{xspace}
\usepackage[T1]{fontenc}
\usepackage{enumerate}
\usepackage[linesnumbered,ruled,vlined]{algorithm2e}
\usepackage{array,multirow}
\usepackage{float}
\usepackage{balance}
\usepackage{tikz}
\usepackage{calc}
\usepackage{subfigure}
\usepackage{listings,amsfonts}
\usepackage{amsmath,pifont}
\usepackage{url}
\usepackage{xspace}
\usepackage{hyperref,endnotes}
\usepackage{array}
\usepackage{multirow,makecell}
\usepackage[misc]{ifsym}
\usepackage{algorithmic}
\usepackage{listings}
\usepackage{svg}
\usepackage{wasysym}

\def\eg{\emph{e.g.}\xspace}

\def\ie{\emph{i.e.}\xspace}
\def\etal{\emph{et al.}\xspace}

\def\BibTeX{{\rm B\kern-.05em{\sc i\kern-.025em b}\kern-.08em
    T\kern-.1667em\lower.7ex\hbox{E}\kern-.125emX}}
    
\begin{document}
\pagestyle{plain}

\title{Cybersquatting in Web3: The Case of NFT}

\author{
\IEEEauthorblockN{Kai Ma\IEEEauthorrefmark{1},
Ningyu He\IEEEauthorrefmark{2}\IEEEauthorrefmark{4},
Jintao Huang\IEEEauthorrefmark{1},
Bosi Zhang\IEEEauthorrefmark{1},
Ping Wu\IEEEauthorrefmark{3},
Haoyu Wang\IEEEauthorrefmark{1}\IEEEauthorrefmark{4}}
\IEEEauthorblockA{\IEEEauthorrefmark{1}Huazhong University of Science and Technology, Wuhan, China}
\IEEEauthorblockA{\IEEEauthorrefmark{2}The Hong Kong Polytechnic University, Hong Kong SAR, China}
\IEEEauthorblockA{\IEEEauthorrefmark{3}Fiberhome Telecommunication Technologies Co., Ltd., Wuhan, China}
\IEEEauthorblockN{\IEEEauthorrefmark{4}Co-corresponding authors: ningyu.he@polyu.edu.hk, haoyuwang@hust.edu.cn}
}

\maketitle

\begin{abstract}
Cybersquatting refers to the practice where attackers register a domain name similar to a legitimate one to confuse users for illegal gains. With the growth of the Non-Fungible Token (NFT) ecosystem, there are indications that cybersquatting tactics have evolved from targeting domain names to NFTs. This paper presents the first in-depth measurement study of NFT cybersquatting. By analyzing over 220K NFT collections with over 150M NFT tokens, we have identified 8,019 cybersquatting NFT collections targeting 654 popular NFT projects. Through systematic analysis, we discover and characterize seven distinct squatting tactics employed by scammers. We further conduct a comprehensive measurement study of these cybersquatting NFT collections, examining their metadata, associated digital asset content, and social media status. Our analysis reveals that these NFT cybersquatting activities have resulted in a significant financial impact, with over 670K victims affected by these scams, leading to a total financial exploitation of \$59.26 million. Our findings demonstrate the urgency to identify and prevent NFT squatting abuses.
\end{abstract}


\section{Introduction}
A Non-Fungible Token (NFT) is a digital identifier that is recorded on a blockchain, which is typically linked to digital assets, like pictures and a piece of music. 
The first NFT was created in May 2014~\cite{nftwiki}. Following that, in 2017, the \texttt{CryptoKitties}~\cite{cryptokitties} drove the popularity of NFTs. 
Since then, with the rise of NFT projects like \texttt{CryptoPunks}~\cite{cryptopunks} and \texttt{Bored Ape Club}~\cite{boredapeyachtclub}, NFTs have become one of the most prominent areas in the cryptocurrency world. 
As of September 2023, there are over 220K different NFT collections, and the total market value of NFTs has surpassed \$10 billion~\cite{nftmarketcap}.

Instead of investors, the prosperity of NFTs has drawn great attention from scammers.
Traditional counterfeit money refers to the creation of imitation fiat currency, such as the US dollar, but without the government's legal sanction~\cite{counterfeit_money}. This old idea has also been reused in cryptocurrency, especially for NFTs. 
Due to the scarcity and exclusivity of NFTs, scammers can counterfeit existing famous NFTs to make a profit.
Such actions carry legal risks and can often constitute illegal activities~\cite{crime_nft,cruz2023cybersquatting}. 

To conduct NFT scams, cybersquatting is a common practice, where attackers register a name similar to a legitimate one to confuse users for illegal gains~\cite{Cybersquatting}.
In recent years, this malicious tactic has evolved beyond traditional domain names and found a new target in the NFT ecosystem.
In NFT marketplaces, each NFT belongs to a collection. Although each collection is uniquely identified by a complex blockchain address, these addresses are difficult to memorize and use. As a result, investors typically rely on collection names when searching for NFTs to trade. This naming mechanism has created an opportunity for scammers to exploit, \textit{i.e.,} they can create NFT collections with names deceptively similar to popular projects, effectively transplanting cybersquatting tactics from domain names to NFT collections to mislead potential investors.

Das~\etal \cite{das2022understanding} has recognized the presence of counterfeit NFTs, but has not delved deeply into this topic. Primarily, they utilized Levenshtein Distance~\cite{levenshtein_distance} to identify potentially similar NFT names. They limit the focus to names exceeding seven characters and a maximum distance of two characters. This method only focuses on syntactic similarity but fails to capture deliberate misspellings and deceptive tactics used in cybersquatting, resulting in a significant number of both false negatives and false positives. For example, a popular NFT collection named \texttt{Azuki}\cite{azukiofficial}, which is shorter than seven characters, and its counterfeit \texttt{Azuki NFT}\cite{azukifake}, with a distance of four characters, will escape detection. Moreover, this approach will also lead to false positives. For instance, the legitimate collection \texttt{CryptoCars}\cite{CryptoCars} was mistakenly flagged as suspicious when compared to another legitimate project, \texttt{CryptoDads}\cite{CryptoDads}, due to a distance of two characters.
Last but not least, their results are limited to pointing out that cybersquatting NFT collections do exist, but no further analysis is conducted, such as systematically analyzing the naming tactics or characterizing the behavior of the participants. 

To bridge this gap, we conducted a thorough and systematic study of such cybersquatting NFT collections.
\textit{First, by adopting cybersquatting in the creation of counterfeit NFTs, our study goes beyond the generic use of \textbf{similarity} measures.} Our detection method, inspired by domain squatting and utilizing squatting tools to design an effective method, is tailored specifically to the unique characteristics of NFT cybersquatting. 
This refined approach allows us to achieve a more comprehensive and accurate detection of counterfeit NFTs.
Unlike previous methods, our approach does not impose length restrictions and has successfully identified seven distinct types of squatting tactics. Our paper represents the first systematic effort to dissect existing squatting patterns—a topic not thoroughly explored by earlier studies. 
\textit{Second, our research provides a more in-depth examination of the counterfeit NFT ecosystem from the perspective of cybersquatting.} 
We have expanded our analysis to encompass not only the naming strategies employed in these scams but also the scale of the operations, scammers and victims, social media, digital asset content and the financial damages incurred. 
This comprehensive view equips stakeholders with a clear and thorough understanding of NFT cybersquatting within the ecosystem.

\textbf{This Work.} In this paper, we present the first comprehensive analysis of NFT cybersquatting on Ethereum. Through analyzing over 220K NFT collections, we have identified 8,019 cybersquatting NFT collections involving over 5.5M tokens that target 654 popular NFT projects, demonstrating the prevalence of NFT squatting in the ecosystem.
Our investigation reveals seven distinct NFT squatting tactics, with combination squatting emerging as the preferred method among scammers, alongside six other mutation squatting tactics. The emergence of these mutation tactics indicates increasing sophistication in squatting strategies (see \S\ref{sec:name_tactic}).
We provide a comprehensive characterization of cybersquatting NFT collections by examining multiple dimensions, including creation time, total supply, market trading activity, active time, associated digital assets, and linked social media presence (see \S\ref{sec:Characterizing_counterfeit_NFT}).
Further analysis centers on key actors in the ecosystem and their financial impact. Our investigation reveals that more than 670K victims have fallen prey to these scams, resulting in significant profits for scammers. The total financial exploitation amounts to approximately \$59.26 million (see \S\ref{sec: actors_profits}).

Our work makes the following contributions.

\begin{itemize}

    \item We provide a systematic analysis of cybersquatting in the Ethereum NFT ecosystem, revealing it as a prevalent issue. This study is the first in-depth examination of this phenomenon. We identified 8,019 NFT squatting projects, with a total of 1,679,896 transfer events, and uncovered seven distinct naming tactics used by scammers.
    \item We identified 794 well-organized scam campaigns, categorized into two types. One type involves malicious links that use NFTs to spread phishing websites, while the other focuses on creating numerous cybersquatting NFTs to maximize potential victims and enhance scam success rates.
    \item We quantified the financial impact of NFT squatting scams, finding that over 670K victims have been affected. Scammers have collectively profited by more than \$59.26 million from these activities.

    \item We constructed a robust database that encompasses a wide range of relevant data, both on-chain and off-chain, which will be shared with the community to aid future research and understanding of NFT cybersquatting.

\end{itemize}

\section{Background}
In this section, we will introduce some necessary background knowledge for this paper.

\vspace{0.05in}
\noindent \textbf{Blockchain \& Ethereum.} Blockchain is a decentralized, immutable ledger managed by a peer-to-peer network, used to securely record transactions. Ethereum is one of the leading open-source blockchain platforms that supports smart contracts that execute when encoded predefined conditions are met.
Deploying and managing digital assets is feasible on Ethereum through smart contracts.

\vspace{0.05in}
\noindent \textbf{Non-fungible Token (NFT).} 
Non-fungible token (NFT) is a type of digital asset verified and distributed on blockchains, distinguished from fungible tokens~\cite{ERC20} by its \textit{irreplaceability} and \textit{indivisibility}.
Typically, each NFT is associated with a URI (Uniform Resource Identifier), pointing to a digital asset that can be accessed by anyone.
To minimize the obstacles to creating and deploying NFTs on Ethereum, two standards are proposed, \ie, \texttt{ERC-721}~\cite{ERC721} and \texttt{ERC-1155}~\cite{ERC1155}. Specifically, by implementing exposed interfaces within these two standards, anyone can create an \textit{NFT collection}. The NFT it contains can be circulated on Ethereum. Their distinctions are that \texttt{ERC-1155} supports both fungible and non-fungible tokens, and also allows for more efficient batch transfers and reduced gas costs.

NFTs can be traded within NFT marketplaces, like \texttt{OpenSea}~\cite{opensea_market}, \texttt{Blur}~\cite{blur_market}, \texttt{LooksRare}~\cite{looksrare_market}, \texttt{X2Y2}~\cite{x2y2_market}, and \texttt{CryptoPunks}~\cite{cryptopunks_market}.
Taking advantage of smart contracts on Ethereum, specifically market contracts, they facilitate the secure and efficient trade of NFTs and offer user-friendly interfaces. 
Moreover, NFT collections leverage social media to promote and attract investors. Platforms like \texttt{Twitter}~\cite{twitter}, \texttt{Discord}~\cite{discord}, and official project websites are widely used to build community engagement and increase visibility~\cite{kapoor2022tweetboost}.

\vspace{0.05in}
\noindent \textbf{NFT Cybersquatting.}
\label{2.5Counterfeit_NFT}
NFT cybersquatting, similar to domain squatting, focuses on exploiting the names of NFT collections to deceive users. Typically, attackers will create NFT collections with identical or confusingly similar names to well-known ones. This enables them to impersonate others, thereby misleading users and capitalizing on the confusion for financial gain.
Attackers employ various tactics to conduct such scams, especially inspired by traditional ERC-20 counterfeit tokens~\cite{gao2020tracking}.
For example, some attackers use \textit{combination squatting techniques}, where additional characters or keywords are appended to the beginning or end of the original name (e.g., \texttt{Azuki\underline{2}}, \texttt{Azuki\underline{NFT})}, creating a deceptive variation.
Another common method is \textit{mutation squatting}, where characters are inserted into or substituted within the middle of the original name (e.g., \texttt{A\underline{h}zuki}, \texttt{AZUK\underline{l}}), making it closely resemble the original while being distinct enough to avoid detection.

\begin{figure}[t]
    \centering
    \includegraphics[width=\columnwidth]{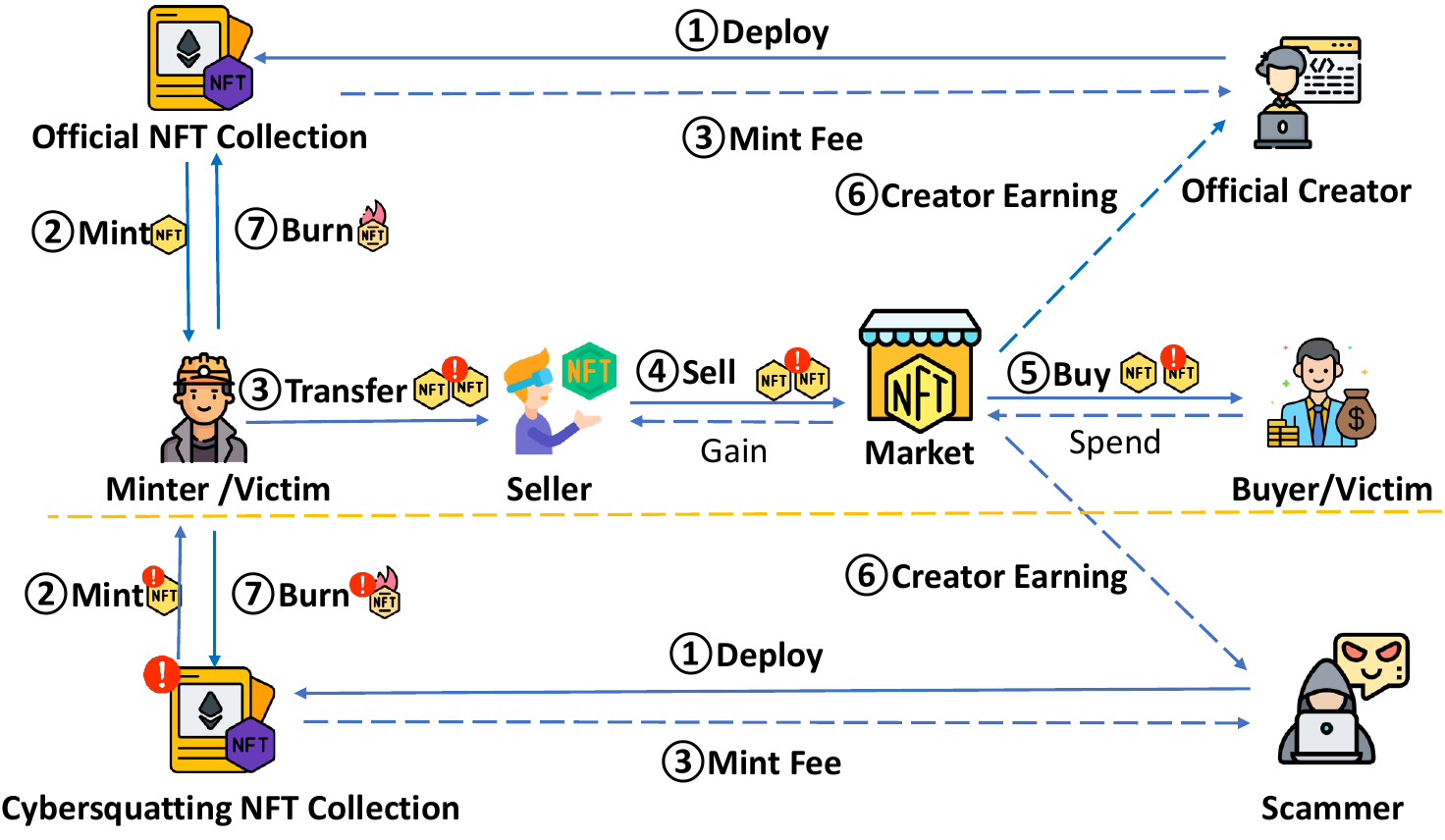}
    \caption{The lifecycle of NFTs, where solid lines represent function calls (\textit{e.g.,} mint and sell), and dashed lines indicate the ETH flow. NFTs are marked with symbols whenever they are transferred between entities.}
    \label{fig:stakeholders}
\end{figure}

\vspace{0.05in}
\noindent \textbf{NFT Lifecycle \& Stakeholders.}
\label{sec:stakeholders}
As depicted in Figure~\ref{fig:stakeholders}, the lifecycle of an NFT can be divided into four key stages: deploying the smart contract, minting, circulating, and burning. The figure highlights both legitimate NFT processes (above the yellow dashed line) and fraudulent cybersquatting NFT processes (below the line). The process begins when a legitimate creator deploys a smart contract on blockchain platforms like Ethereum (Step 1). Users are then invited to mint NFTs (Step 2) by paying a piece of mint fee~\cite{huang2023miracle} (Step 3), which is subsequently transferred to the creator's account. 
After minting, the NFT enters circulation, where it can be sold (Step 4) or bought (Step 5) on secondary markets. Part of the received money is transferred to the creator as creator earnings, a percentage of which is determined by the royalty percentage explicitly set by the creator~\cite{nft_royalties} (Step 6). 
Eventually, some NFTs may be removed from circulation through a process called burning (Step 7).
Scammers can exploit this system by deploying cybersquatting NFT contracts. If they are not detected by secondary markets, the corresponding NFT will be circulated as if it were the original one.
Note that both mint fees and creator earnings are profits for the creator of NFT collections. Thus, scammers will attract users to mint these cybersquatting NFTs or trick users into buying and selling these NFTs in the secondary market to make a profit.

We define victims as entities that have direct financial contributions to scammers. It can be achieved through two channels: (1) participants who pay minting fees to mint NFTs (i.e., minters), and (2) secondary market buyers who purchase creator earnings (i.e., buyers).

\section{Study Design}
\label{sec:study_design}
In this section, we list all research questions regarding NFT cybersquatting, the data we collected, and the adopted methodology to identify NFT cybersquatting.

\subsection{Research Questions}
Regarding the NFT cybersquatting phenomenon, our study is driven by the following research questions (RQs):

\begin{itemize}

\item[RQ1] \textbf{What are the naming tactics used by scammers for creating cybersquatting NFT collections?}
To answer this research question, we first categorize existing cybersquatting naming tactics into three categories. Then, we characterize their distributions and longitudinal evaluations.

\item[RQ2] \textbf{What are the characteristics of cybersquatting NFT collections?}
This question focuses on analyzing the characteristics of cybersquatting NFT collections, including their metadata (like creation time, total supply, and active periods), related digital assets, and linked social media accounts.
Answering this research question helps the community to better understand the distinct traits of these collections.

\item[RQ3] \textbf{Who are the actors behind NFT cybersquatting? What are the profits gained?}
This question focuses on thoroughly characterizing the actors involved in NFT cybersquatting, namely the scammers and victims, and exploring the potential existence of organized scam groups operating at scale. Moreover, this question also aims to quantify the profits obtained by scammers through mint fees and creator earnings, shedding light on the financial incentives driving these scams.
\end{itemize}

\subsection{Data Collection}
\label{sec:datacollect}
To comprehensively investigate the NFT cybersquatting phenomenon, based on several previous studies~\cite{huang2023miracle,das2022understanding}, we have gathered a diverse dataset, as outlined in Table~\ref{tab:data}.
Specifically, we have collected \textit{Smart Contract}, \textit{Transfer Event}, \textit{Market Trade}, and \textit{NFT Metadata}. The data from the first three categories is retrieved on-chain directly.
As for \textit{NFT Metadata}, including image URI and social media information, we utilized APIs from prominent platforms such as \texttt{OpenSea}~\cite{opensea_2022_opensea} and \texttt{ChainBase}~\cite{chainbaseAPI}.

\begin{table}[t]
\centering
\caption{The statistics of our dataset (till Sep. 1st, 2023).}
\label{tab:data}
\resizebox{\columnwidth}{!}{%
\begin{tabular}{@{}rccc@{}}
\toprule
\multicolumn{1}{l}{\textbf{Items}}  & \textbf{ERC-721} & \textbf{ERC-1155} & \textbf{Total} \\ \midrule
\multicolumn{1}{l}{\textbf{Smart Contract}}        & 186,111         & 34,807           & 220,918 \\
\multicolumn{1}{l}{\textbf{Transfer Event}}  & 219,016,571     & 26,361,227        & 245,377,798\\
\textit{mint}                                & 141,921,955     & 8,454,808         & 150,376,763 \\
\textit{burn}                                & 3,083,460      & 1,346,984            & 4,430,444  \\
\textit{swap}                                & 74,011,156      & 16,559,435       & 90,570,591 \\
\multicolumn{1}{l}{\textbf{Market Trade}\ \ \ \ \ \ \ \ \ \ \ }    & \multicolumn{2}{c}{98,390,236}                & - \\
\textit{OpenSea}                     & \multicolumn{2}{c}{93,484,505}        & - \\
\textit{LooksRare}                           & \multicolumn{2}{c}{620,789}          & - \\
\textit{X2Y2}                                & \multicolumn{2}{c}{2,270,023}        & - \\
\textit{Blur}                                & \multicolumn{2}{c}{1,983,974}        & - \\
\textit{CryptoPunks}                                & \multicolumn{2}{c}{30,945}        & - \\

\multicolumn{1}{l}{\textbf{NFT Metadata}\ \ \ \ \ \ \ \ }  & \multicolumn{2}{c}{220,918} &-\\
\textit{Twitter Username}                                 &35,774         &5,524           &41,298 \\
\textit{External Link}                                &47,972       &9,091        &57,063  \\
\textit{Image URI}                                & 76,632,735              & 4,091,076                & 80,724,191 \\

\bottomrule
\end{tabular}
}
\end{table}

\subsubsection{Smart Contract \& Transfer Event}
\label{sec:transfer_event}
To study cybersquatting NFTs as comprehensively as possible, we first need a complete list of NFT projects. 
Within both the \texttt{ERC-721} and \texttt{ERC-1155} standards, NFT transfers emit specific types of events that will be recorded on-chain. Thus, we have deployed a client node \texttt{Geth}~\cite{geth} and synchronized all blocks up until September 1st, 2023. 
For \texttt{ERC-721}, the topic signature of the \texttt{Transfer} event is \texttt{0xddf252ad}. For \texttt{ERC-1155}, the \texttt{TransferSingle} and \texttt{TransferBatch} events correspond to \texttt{0xc3d58168} and \texttt{0x4a39dc06}, respectively, to represent single and batch NFT transfers.
In total, we have gathered 219,114,287 \texttt{ERC-721} and 27,548,181 \texttt{ERC-1155} transfer events.
Based on them, we have parsed 220,918 unique involved contract addresses, further categorized into 186,111 \texttt{ERC-721} and 34,807 \texttt{ERC-1155} smart contracts based on their distinct function signatures. 
Additionally, these transfer events can be categorized into three types: \textit{mint}, \textit{burn}, and \textit{swap}, where mint events have a null address in the from field, burn events have a dead address in the to field, and all other transfers are classified as swaps, representing ownership changes.

\subsubsection{Market Trade}
A market trade is a transaction within a secondary marketplace smart contract involving a monetary exchange, unlike a transfer, which only changes the ownership of NFTs. Following the method from Huang et al.~\cite{huang2024unveiling}, we collected all trade records from all five secondary market contracts.
Each time an NFT is sold, these marketplaces emit specific trade events, such as the \texttt{TakerAsk} event on \texttt{LooksRare} or the \texttt{PunkBought} event on \texttt{CryptoPunks}. By parsing these events, we can extract some key information, including the transaction hash, seller, buyer, and price.
By September 1st, 2023, we had gathered 97,902,053 trade records from five major secondary markets.

\subsubsection{NFT Metadata}
When an NFT creator or an authorized representative lists an NFT collection on a secondary marketplace, they are required to provide some metadata about the collection, including the NFT name, creator, social media links, and royalty percentage.
Using the APIs provided by \texttt{OpenSea} and \texttt{ChainBase}, we tried to extract the metadata for all 220K collected NFT collections, specifically focusing on \texttt{Twitter usernames}, \texttt{external links}, and \texttt{image URIs}.
In total, we gathered 41,298 unique \texttt{Twitter Usernames} and 57,063 distinct \texttt{External Links}. Additionally, we retrieved 80,724,191 \texttt{Image URLs} corresponding to individual NFTs.
We made our best efforts to download as many images as possible from these links and successfully collected 10,761,760 images, providing a valuable dataset for further analysis of NFT collections.

\begin{figure*}[t]
    \centering
    \includegraphics[width=0.9\textwidth]{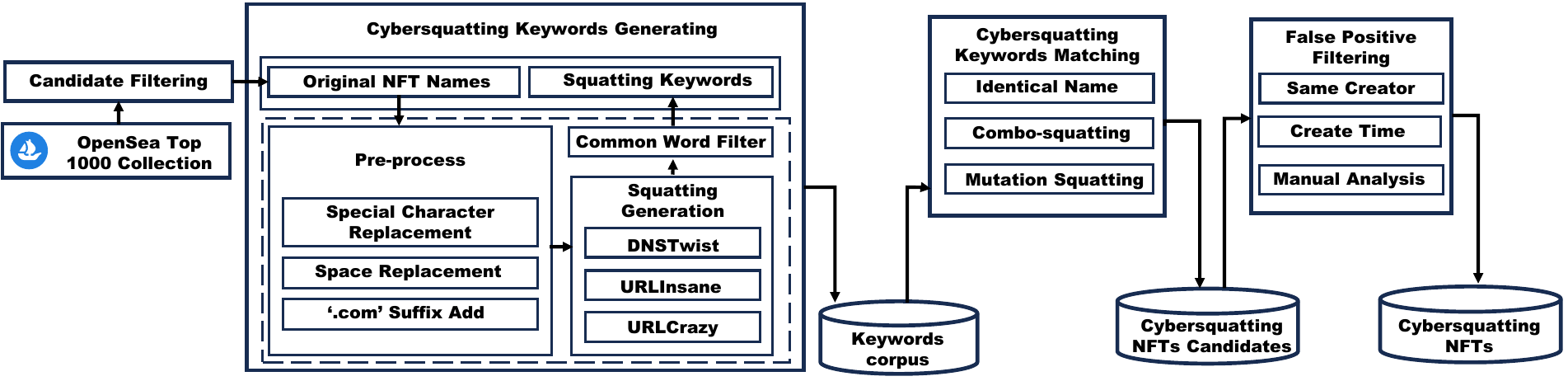}
    \caption{The workflow of identifying cybersquatting NFT collections. }
    \label{fig:nftworkflow}
\end{figure*}

\subsection{Identifying NFT Cybersquatting}
\label{sec:3.3}
We have developed a three-stage NFT cybersquatting detection method, as illustrated in Figure~\ref{fig:nftworkflow}. In general, the process takes 220K collected NFT collection names as inputs, and outputs all cybersquatting NFT collections.

\subsubsection{Stage I: Candidate Filtering}
Similar to the cybersquatting phenomenon observed in ERC-20 tokens~\cite{gao2020tracking}, attackers tend to target NFT collections that exhibit high user engagement and trading volume. Thus, we curated a list of the top 1,000 NFT collections based on their total market capitalization, using data sourced from OpenSea.
We found that some large projects may release multiple NFT collections, whose collections may share an identical name. To avoid duplication, we only keep the most influential collection, i.e., with the highest market cap, for each project.
For example, both \#52 and \#372 from the \texttt{Decentraland} project share the same project name but represent different NFT collections.
We finalized a list of 996 distinct NFT collections to serve as candidates for further analysis.

\subsubsection{Stage II: Cybersquatting Keywords Generating}
Based on 996 names collected in Stage I, we decided to apply current squatting methods to synthesize a keyword corpus, and examine if the name of any non-popular NFT collections hits a keyword.
Specifically, due to the lack of NFT-specific squatting tools, we adopt \texttt{URLCrazy}~\cite{urlcrazy}, \texttt{URLInsane}~\cite{urlinsane}, and \texttt{DNSTwist}~\cite{dnstwist}, which are originally designed for domain squatting.
Thus, we pre-processed the 996 names by replacing special characters with spaces and converting spaces to dots to form domain-like strings.
After the generation, we removed unnecessary suffixes and restored them to valid NFT collection names.
Note that before constructing the final keyword corpus, we have to remove some of the generated names that are common words. For instance, for the \texttt{Metaverse HQ} collection~\cite{metaversehq}, the generated name may be \texttt{Metaverse} via the omission squatting strategy. Since such a common word may be included by many benign NFT collections, we use a public English~\cite{CommonEnglishWords} and cryptocurrency-specific common word list~\cite{Cryptocurrency_common_world} to refine the generated names.

\subsubsection{Stage III: Cybersquatting  Keywords Matching}
As for the matching process, it can be divided into two steps. First, we performed an exact match to identify cybersquatting NFTs out of the 220K NFT collections. Then, we conducted a partial match using the keywords from the corpus to identify potential cybersquatting NFTs, \ie, the generated name could be part of the NFT collection name, while ignoring cases and special characters.

\subsubsection{Stage IV: False Positives Filtering}
\label{sec:3.3.4}

False positives may arise from legitimate derivative or test NFTs with similar names or from trustworthy creators whose collections resemble official projects. To mitigate these issues, we applied filtering rules inspired by Hu et al.~\cite{hu2020mobile} and Gao et al.~\cite{gao2020tracking}, as well as traditional domain squatting detection methods~\cite{szurdi2014long}, which have demonstrated high-precision detection through rigorous validation. 
A multi-step verification approach is implemented for a ground truth-like accuracy in identifying cybersquatting NFTs.  

First, we filtered out derivative NFT collections officially created by the same team with similar names, such as \texttt{Doodles} and \texttt{Space Doodles}. 
Then, we excluded NFTs deployed before the official ones, ensuring that legitimate projects were not misclassified.  
Last, to further refine our results, we incorporated a manual review process where flagged NFTs were evaluated based on multiple factors, including price trends, transfer activity, social media status, external labels, and image similarity. 
Specifically,
a collection is labeled as suspicious if it meets at least four of the following criteria: (1) the floor price drops by more than 90\% from its peak and remains unrecovered for 30 days; (2) the number of monthly transfers falls by more than 90\% from its peak and stays low for two consecutive months; (3) no social media activity, such as Twitter posts, is observed within 30 days after the last on-chain activity; (4) the contract is explicitly marked as malicious on public sources like Etherscan or Chainabuse; (5) the collection’s images are visually similar to those of the target collection, with DHash distances below 5. 
These thresholds are selected based on prior work~\cite{huang2023miracle,das2022understanding}. This design ensures reproducibility and enables automated implementation at scale.

We emphasize that the last-step manual review process was conducted solely to ensure high precision during evaluation, but \textit{it can also be fully automated}.
For example, price drops and transfer inactivity can be identified using time-series thresholds (e.g., over 90\% within 30 days), and the social media silence can be verified via Twitter API.
These detectors can be integrated into a scalable pipeline that flags suspicious collections based on a majority rule, \textit{e.g.,} four out of five are triggered. 
This architecture supports real-time detection and can be further optimized using supervised learning models trained on labeled datasets.

\section{RQ1: NFT Cybersquatting Naming Tactics}
\label{sec:name_tactic}
This section explores the naming tactics used in NFT cybersquatting, highlighting those ones such as identical name replication, combination squatting, and mutation-based squatting that enable scammers to deceive users.

\subsection{Overall Result}
We have identified 8,019 cybersquatting NFT collections, including 6,495 \texttt{ERC-721} and 1,525 \texttt{ERC-1155} ones, involving 1,679,896 transfer events and 780,550 unique Ethereum addresses. 
These cybersquatting NFT collections target 654 of the 996 popular NFT collections we studied (65.66\%). 
Table~\ref{tab:top5target} lists the top-5 official NFT collections with the highest number of cybersquatting instances, where the rank is calculated by their market capitalization.
As we can see, the collection \texttt{mfers}~\cite{mfers} has the most cybersquatting NFT collections, with a total of 552. Following closely, \texttt{Doodles}~\cite{doodles} has 508 cybersquatting instances. In total, these five collections have been cybersquatted 1,850 times.
Additionally, one of the most well-known NFT collections, Bored Ape Yacht Club, also has 108 cybersquatting cases, indicating that high-value projects are often taken as cybersquatting targets.

\textbf{\textit{Findings:}} Our analysis reveals that NFT cybersquatting tends to disproportionately target collections with higher market capitalizations. Specifically, 45\% of all cybersquatting collections target the Top 100 NFT collections by market cap, and this percentage rises to 78\% when considering the Top 500 collections. This aligns with the intuition that collections with higher market caps are more likely to be targeted by cybersquatting schemes,

\begin{table*}[t]
\centering
\caption{Top-5 cybersquatted NFT collections, where the rank is obtained by the corresponding market cap.}
\label{tab:top5target}
\begin{tabular}{@{}cclc@{}}
\toprule
\textbf{Rank} & \textbf{Name} & \textbf{Contract Address}                  & \textbf{Time} \\ \midrule
28   & mfer          & 0x79fcdef22feed20eddacbb2587640e45491b757f & 552           \\
9    & Doodles       & 0x8a90cab2b38dba80c64b7734e58ee1db38b8992e & 508           \\
4    & Azuki         & 0xed5af388653567af2f388e6224dc7c4b3241c544 & 334           \\
8    & Moonbirds     & 0x23581767a106ae21c074b2276d25e5c3e136a68b & 283           \\
128  & Nouns         & 0x9c8ff314c9bc7f6e59a9d9225fb22946427edc03 & 173           \\ \bottomrule
\end{tabular}%
\end{table*}

\subsection{NFT Cyberquatting Naming Tactics}
\label{sec:nametactics}
For the identified 8,019 cybersquatting NFT collections, we focus on analyzing the various naming tactics employed by scammers to create deceptive collections. In this section, we provide a detailed characterization of these tactics. To achieve this, we first conduct a thorough review of all mutation strategies utilized by popular tools such as \texttt{DNSTwist}, \texttt{URLCrazy}, and \texttt{URLInsane}, which are commonly used to generate variations of legitimate names for phishing and cybersquatting attacks. Based on this review, we reclassify or merge the tactics according to their actual effects and how effectively they mimic legitimate NFT collections.
In addition, we also refer to the classification methodology proposed by~\cite{hu2020mobile} to refine our categories. By combining these insights with our own analysis, we establish a robust and comprehensive classification of naming tactics employed in NFT cybersquatting.

\subsubsection{Overall Result}
Ultimately, for the 8,019 identified collections, seven distinct naming tactics are observed, as outlined in Table~\ref{tab:summaryofsquatting}.
Among these tactics, directly adopting an identical name accounts for only 8.77\% of all cases. While this method may be the most intuitive and can yield highly deceptive results, the identical name also makes these counterfeits easier to detect and flag by users and platforms.
As a result, most cybersquatting NFT collections have opted for more sophisticated naming tactics, such as \textit{combination squatting} and \textit{mutation-based squatting}, which allow counterfeit collections to appear more distinct yet still similar enough to deceive users. These advanced tactics are discussed in detail in the following sections.

\noindent \textbf{Compare with SOTA.}  
Das et al.~\cite{das2022understanding} adopt a Levenshtein Distance based approach to detect NFT squatting. However, its approach requires that NFT names must be longer than seven characters and limits the maximum edit distance to two, leaving 158 of the top 1,000 NFT collections undetectable.
Moreover, when applying their method to our dataset, it can only identify 424 out of 7,316 non-identical cybersquatting cases, missing nearly 7,000 instances. 
We have manually rechecked 111 marked cases that were not present in our dataset. We figured out that 85\% (94 cases) were not cybersquatting attempts. Their names indeed met the Levenshtein Distance conditions but the corresponding project did not intend to imitate popular NFTs. For example, \texttt{CryptoDads} was mis-bound with \texttt{CryptoGods}, \texttt{CryptoDates}, and \texttt{CryptoCars}, all of which are independent projects. The remaining 15\% (17 cases) were legitimate cybersquatting attempts that our method did not capture, such as \texttt{y00ts Yacht Club} and \texttt{r00ts Yacht Club}.
This is because the substitution like \texttt{y} to \texttt{r} falls outside the scope of our mutation patterns. We underline that our method mainly considers keyboard adjacency or visually similar character substitutions (e.g., \texttt{y} to \texttt{u} on a QWERTY keyboard or \texttt{i} to \texttt{l}).
This analysis shows that \textit{Levenshtein Distance only focuses on syntactic similarity but fails to capture deliberate misspellings and deceptive tactics used in cybersquatting}. Our approach, inspired by traditional domain squatting techniques, accounts for a wider range of tactics, offering more accurate and comprehensive detection. 
Additionally, our study expands beyond name similarity, examining seven distinct naming tactics, attacker behaviors, victim impacts, and financial losses, offering a more holistic understanding of NFT cybersquatting.

\subsubsection{Combination Squatting}
Combination squatting refers to adding extra characters or words before or after an existing official NFT collection name.
This tactic accounts for the majority (5,391, 67.22\%) of the NFT cybersquatting cases.
Unlike mutation-based squatting, combination squatting retains the original spelling of the official name.
We further investigated the most commonly used keywords in combination squatting for NFTs.
The top five added keywords are \textit{the} (421), \textit{NFT} (197), \textit{official} (171), \textit{by} (142) and \textit{collection} (122). 
For example, \texttt{Lives of Asuna} is altered to \texttt{The Lives of Asuna} or \texttt{Lives of Asuna NFT}. 
According to our statistics, the transactions of such cybersquatting NFTs account for 74.79\% of the total, and the addresses involved represent 83.53\% of the total. This suggests that combination squatting is both the most popular and effective strategy. It allows those cybersquatting collections to masquerade as the derivative series of official collections, often going undetected by third-party platforms~\cite{opensea_market}, thereby providing a significant window of opportunity for counterfeiters.

\begin{table*}[t]
\centering
\caption{The distribution of naming tactics of cybersquatting NFT collections.}
\label{tab:summaryofsquatting}
\begin{tabular}{llccc}
\toprule
\multicolumn{2}{l}{\textbf{Naming Tactics}}                                                                                                                                                     & \textbf{ERC-721} & \textbf{ERC-1155} & \textbf{Total}   \\ \midrule
\multicolumn{2}{l}{\textbf{Identical Name}}                                                                                                                                           & 654              & 49                & 703 (8.76\%)    \\ \midrule
\multicolumn{2}{l}{\textbf{Combination Squatting}}                                                                                                                                    & 4,217            & 1,174             & 5,391 (67.22\%) \\ \midrule
\multirow{6}{*}{\textbf{\begin{tabular}[c]{@{}l@{}}Mutation-based \\ Squatting\end{tabular}}} 
                                                                                              & \textit{Character Insertion}                                                          & 70               & 45                & 115 (1.43\%)    \\
                                                                                              & \textit{Character Omission}                                                           & 927              & 177               & 1,104 (13.76\%) \\
                                                                                              & \textit{Case Substitution}                                                            & 385              & 72                & 457 (5.69\%)    \\
                                                                                              & \textit{\begin{tabular}[c]{@{}l@{}}Misspelling Mistakes \\ Substitution\end{tabular}} & 223              & 8                 & 231 (2.88\%)    \\
                                                                                              & \textit{Homoglyph}                                                                    & 16               & 0                 & 16 (0.19\%)     \\
                                                                                              & \textit{Homophone}                                                                    & 2                & 0                 & 2 (0.02\%)      \\ \midrule
\multicolumn{2}{l}{\textbf{Total}}                                                                                                                                                    & 6,494            & 1,525             & 8,019            \\ \bottomrule
\end{tabular}%
\end{table*}

\subsubsection{Mutation-based Squatting}
\label{sec:mutation}
Mutation-based squatting refers to the tactic where scammers alter an official NFT name by inserting, removing, or substituting characters to create a deceptive variation that closely resembles the original.
Though dozens of types of squatting names can be generated, we only observe the following six kinds of mutation-based tactics in real-world:
\begin{enumerate}
    
    \item \textit{Character Insertion:} Inserting a meaningless character, \eg,"Moonbirds" into "Moonb\underline{h}irds".

    \item \textit{Character Omission:} Deleting one or more characters, \eg, "Doodle\underline{s}" into "Doodle".

    \item  \textit{Case Substitution:} 
    Replace a lowercase letter with an uppercase letter, or vice versa, \eg, "M\underline{i}lady Maker" into "M\underline{I}lady Maker". 
    
    \item  \textit{Misspelling Mistakes Substitution:} Replacing a vowel character with another vowel character, \eg, "M\underline{i}lady Maker" into "M\underline{a}lady Maker". 
    
    \item \textit{Homoglyph:} Replacing a character with another character with similar appearance, \eg, "Azuk\underline{i}" into "AZUK\underline{l}".
    
    \item \textit{Homophone:} Replacing a word with another word with the same pronunciation, but differing in meaning, \eg, "\underline{Bored} Ape Yacht Club" into "\underline{Board} Ape Yacht Club".
\end{enumerate}

Among them, character omission is adopted the most widely, where its number has reached 1,104 (13.76\%).
The official NFT \texttt{Doodles} is the biggest victim of this strategy, with 235 cybersquatting collections omitting the \texttt{s} from \texttt{Doodles}
Moreover, \textit{case substitution} and \textit{misspelling mistakes substitution} are also common, accounting for 5.69\% and 2.88\%, respectively, which only introduce minor changes that easily deceive users.
These three strategies nearly make up all mutation-based squatting cases, requiring additional caution.

\textbf{\textit{Findings:}} We have identified 8,019 cybersquatting NFT collections, with only 8.77\% using direct name copying. Attackers tend to employ more sophisticated strategies, like combination squatting and mutation-based squatting, to avoid being detected.

\subsubsection{Tactics Evolution}
\begin{figure}
    \centering
    \includegraphics[width=0.9\linewidth]{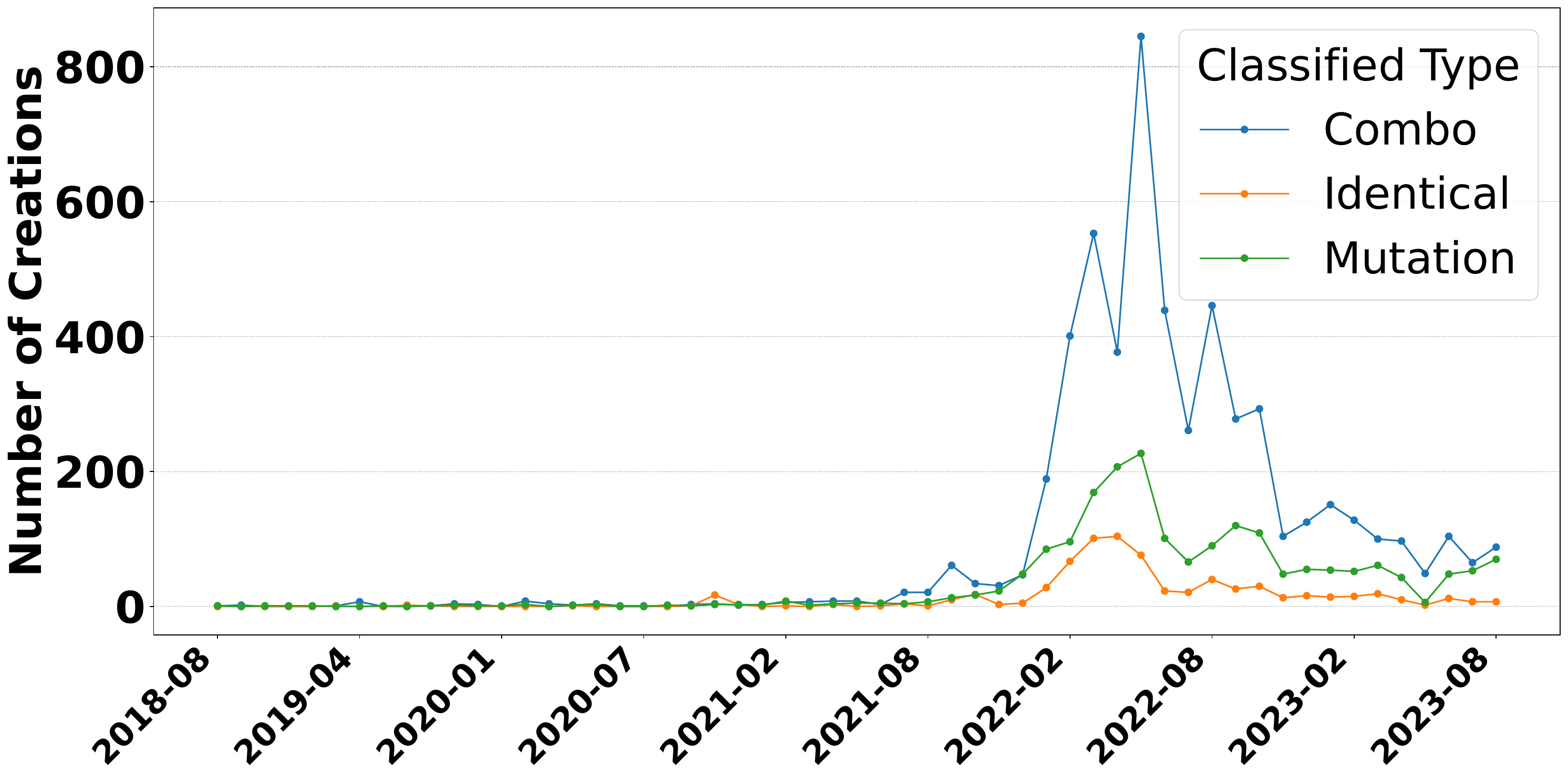}
    \caption{The longitudinal evolution of naming tactics of cybersquatting NFT collections.}
    \label{fig:name_time}
\end{figure}

Figure~\ref{fig:name_time} depicts the evolution of naming tactics used in NFT cybersquatting over time.
The most prominent naming tactic is \textit{combination squatting}, which has consistently been the dominant approach, peaking at 845 creations in June 2022.
We can also observe the steady presence of \textit{mutation-based squatting}, with notable spikes corresponding to the overall growth of NFT markets. 
As we illustrated in \S\ref{sec:mutation}, this technique allows scammers to make slight changes to the collection names, making detection by automated systems more difficult while maintaining enough resemblance to legitimate collections to fool users.
Finally, the least-used tactic is \textit{identical name}. Although effective in some cases, this method is easily flagged by platforms or users, leading to its lower adoption. As the figure shows, identical name squatting has remained consistently low over time, further demonstrating that scammers prefer tactics that involve subtle variations to avoid detection.

\noindent \textbf{Comparison with Other Domain-Specific Squatting.} 
Naming strategies in cybersquatting attacks have been widely studied across different domains, including traditional domain squatting, ERC-20 token squatting, and mobile App squatting. Though the ecosystem varies, common patterns consistently emerge in how attackers manipulate names to deceive users and evade detection efforts.
Specifically, in traditional domain squatting research, Zeng~\etal~\cite{zeng2019comprehensive} found that combination squatting in the domain attracted six times more traffic than all other squatting techniques combined, making it the most effective method overall. 
In ERC-20 squatting research, Gao~\etal~\cite{gao2020tracking} studied counterfeit ERC-20 tokens and found that approximately 77\% were created through combination squatting, while the remaining 23\% used identical names\footnote{They did not consider mutation-based squatting.}. Regarding mobile App squatting, Hu~\etal~\cite{hu2020mobile} found that among all squatting strategies, combination squatting had the highest single-category proportion (30\%), followed by case substitution (17\%), which remains a notable technique for obfuscating app identities. 

In the case of NFT cybersquatting, according to our results, attackers lean heavily on combination squatting (67.22\%), aligning with trends observed in all domain squatting, ERC-20 squatting, and App squatting.
Its popularity is that it allows fraudulent collections to mimic official projects while evading immediate detection. 
However, unlike ERC-20 squatting, where identical names were once common, there are only 8.76\% identical name squatting in the NFT ecosystem. We speculate the reasons are twofold. On one hand, users' awareness of counterfeit tokens has improved, making direct replication less effective. On the other hand, NFT marketplaces enforce stricter name uniqueness policies. Consequently, attackers in the NFT ecosystem have increasingly turned to mutation-based squatting (24\%), using tactics such as character omission (13.76\%), case substitution (5.69\%), and misspelling substitution (2.88\%) to create deceptive variations that bypass automated detection. This shift towards mutation-based tactics allows fraudsters to circumvent detection systems by making subtle changes to the names, while still keeping them highly recognizable to users.
These findings highlight that \textit{NFT cybersquatting tactics align with established squatting patterns in other ecosystems but also evolve in response to platform defenses and user behavior}. While combination squatting remains the preferred strategy, mutation-based squatting serves as a crucial adaptation to stricter enforcement and improved scam awareness in the NFT marketplace.

\textbf{\textit{Findings:}} 
While combination squatting remains the dominant approach due to its high effectiveness, the increasing prevalence of mutation-based squatting reflects a shift toward more subtle tactics that better balance deception and evasion.

\begin{framed}
\noindent\textbf{Answer to RQ1:}
\textit{
NFT cybersquatting scams are prevalent, where over 60\% of top NFT collections are targeted and we have uncovered 8,019 cybersquatting ones.
Like ERC-20 counterfeit, combination squatting is the most common and effective method favored by scammers. However, the more sophisticated mutation-based squatting also plays a significant role in deceiving users and cannot be overlooked in the NFT ecosystem. 
}
\end{framed}

\section{RQ2: Characterizing Cybersquatting NFT Collections}
\label{sec:Characterizing_counterfeit_NFT}
In this section, we try to characterize some basic information about cybersquatting NFT collections, including their metadata, related digital assets, the linked social media accounts, and external links.

\subsection{Basic Feature}
\label{sec:basic_feature}
\begin{figure}
    \centering
    \includegraphics[width=0.8\columnwidth]{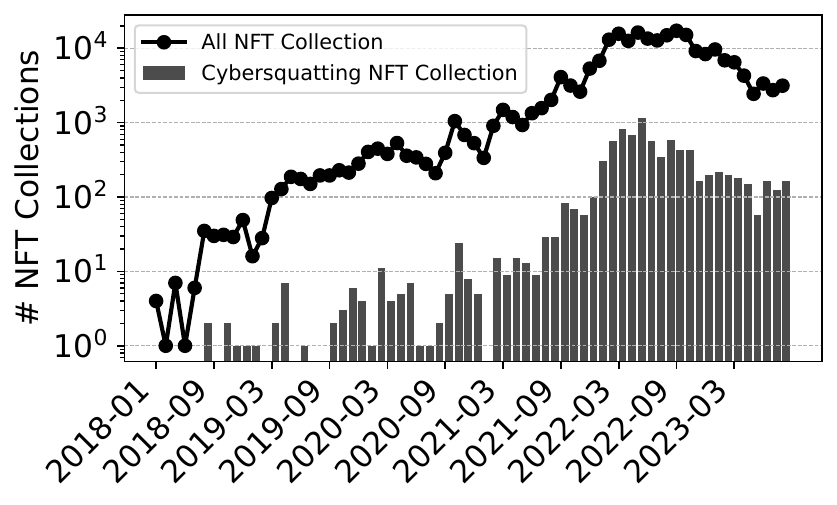}
    \caption{The distribution of (cybersquatting) NFT collections according to the deployment time.}
    \label{fig:CounterfitTime}
\end{figure}

\subsubsection{Creation Time}
Figure~\ref{fig:CounterfitTime} shows the distribution of cybersquatting NFT collections according to their deployment time. Since September 2021, the number of cybersquatting NFTs has steadily increased, peaking at 1.1K in June 2022. This trend closely mirrors the growth of the overall NFT market, indicating that scammers are taking advantage of the rising popularity of NFTs to generate cybersquatting collections and exploit naive buyers.

\subsubsection{Total Supply}
\begin{figure}
    \centering
    \includegraphics[width=0.9\columnwidth]{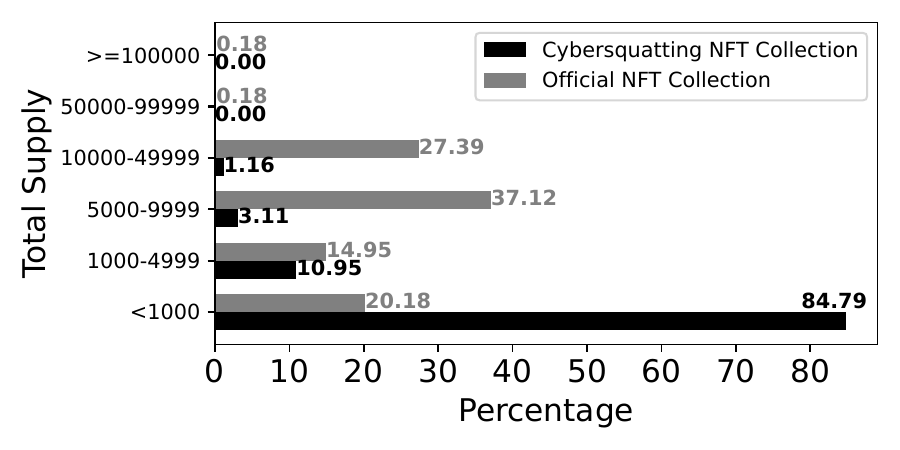}
    \caption{The comparison of the total supply between official and cybersquatting NFT collections.}
    \label{fig:totalsupply}
\end{figure}

In the NFT ecosystem, total supply is a key factor that influences the price of NFTs. A lower total supply often creates an illusion of scarcity, driving up demand as potential buyers rush to acquire an NFT before the limited stock is depleted. 
We examined the total supply of these cybersquatting NFT collections, as shown in Figure~\ref{fig:totalsupply}.
More than 84\% cybersquatting NFT collections mint less than 1,000 NFTs, while nearly 65\% of legitimate NFT collections tend to mint more than 5,000 NFTs.
We suspect there are three reasons.
First, it reduces costs associated with minting, particularly the handling of content and gas fees.
Second, a lower supply creates a sense of scarcity, encouraging buyers to act quickly to avoid missing out. 
Finally, reducing the total supply minimizes exposure. Larger supplies are more likely to attract market attention, increasing the likelihood of being detected and removed by platforms. 

\subsubsection{Market Trade} 
\begin{figure}
    \centering
    \includegraphics[width=0.9\columnwidth]{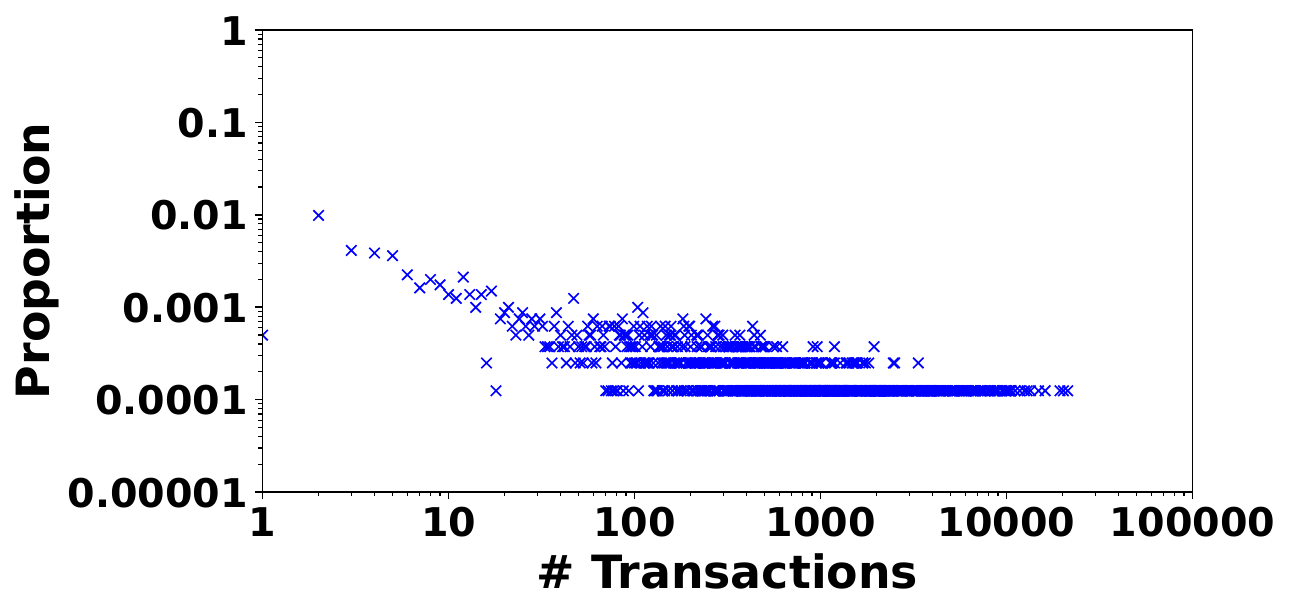}
    \caption{The distribution of cybersquatting NFT collections according to the number of transactions.}
    \label{fig:Proportion_Tran}
\end{figure}

We further characterize the number of trade records of the identified cybersquatting NFT collections in secondary markets.
Figure~\ref{fig:Proportion_Tran} illustrates the distribution. According to our statistics, over 80\% of them have no more than 100 trades, and only 5.17\% of them have more than 1,000 trades. 
Additionally, we found that approximately 90\% of the trades occurred on OpenSea, with the remaining 10\% distributed among the other four marketplaces. This distribution is largely due to OpenSea's prominence as the largest and most popular secondary market for NFTs, attracting a majority of trading activity.
The low number of trades can be attributed to the short lifespan of these cybersquatting NFTs, whose primary goal is obtaining quick profits before being detected. As a result, scammers do not invest in long-term engagement or marketing efforts, leading to fewer trades overall.

\subsubsection{Active Time}
\begin{figure}
    \centering
    \includegraphics[width=0.8\columnwidth]{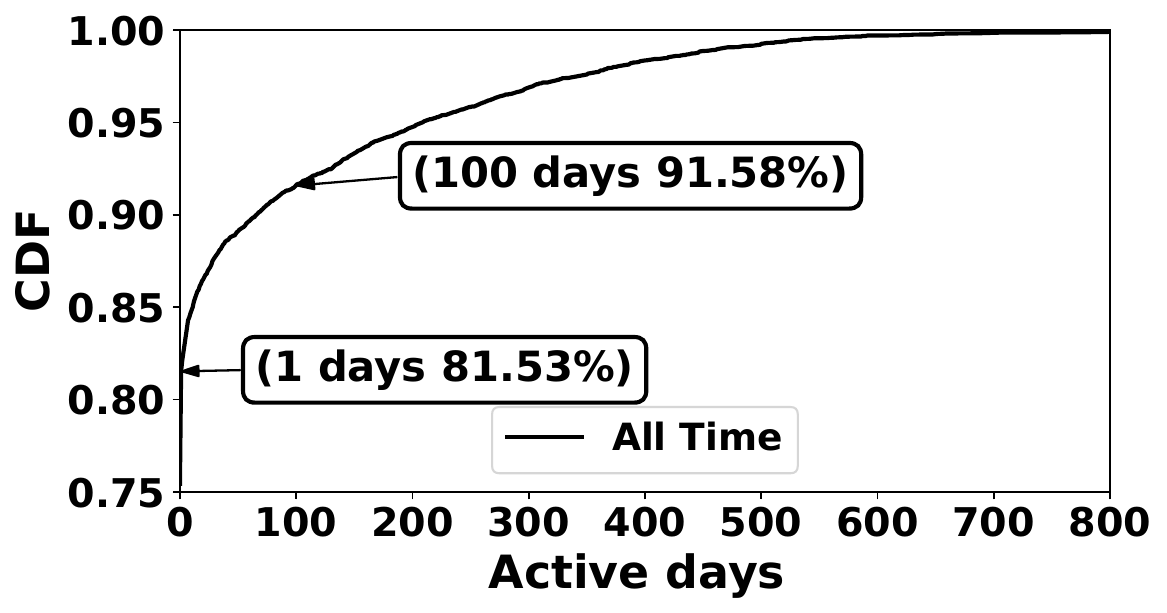}
    \caption{The CDF of cybersquatting NFT collections according to their active period.}
    \label{fig:CDF}
\end{figure}

Active time is defined as the period between the first and the last transaction in secondary markets.
Figure~\ref{fig:CDF} shows the active period for all identified cybersquatting NFT collections.
As we can see, over 80\% of them are only active for no more than a single day, and only 8.41\% of them are active over 100 days.
These findings lead us to speculate that the primary objective of these cybersquatting NFT collections is to earn a quick profit. Consequently, most of them exhibit a relatively brief period of activity, coupled with a lower number of transactions. This enables scammers to swiftly move on to new deceptive schemes once their initial goals are achieved. Thus, the short-lived nature and lower transaction volumes of these fraudulent NFTs reflect their underlying profit-driven motivations.

\textbf{\textit{Findings:}}
Our analysis reveals that cybersquatting NFTs are designed for quick profits with minimal investment. These cybersquatting NFT collections typically have lower total supply, shorter lifespans, and limited trading activity.
Scammers focus on exploiting the market quickly before being detected, reflecting the short-term, profit-driven nature of these fraudulent schemes.

\subsection{Content Theft}
\label{sec:content_theft}
The content linked to by an NFT URI is critical to its uniqueness and value, as it typically represents the digital asset associated with the token.
Therefore, we further examine the content of NFTs in the cybersquatting collections in comparison to their corresponding official ones to determine if \textit{content theft} has occurred, \ie, unauthorized use or duplicated content.
More specifically, we identify two main types of content theft, \ie, \textit{URI theft}, which involves directly copying the official URI, and \textit{image theft}, where the official content is stolen, but the URI is altered.

\subsubsection{URI Theft} To make malicious projects appear more like well-known ones, some cybersquatting NFT collections simply copy the URIs from popular collections. To assess the scale of this issue, we compared the URIs of NFTs in popular collections with those in identified cybersquatting collections. Our analysis revealed that 64 official NFT collections had their URIs stolen by 77 cybersquatting collections, involving a total of 495,418 pairs of identical NFT URIs. In the 64 official NFT collections with stolen URIs, only 11 collections were targeted by two separate cybersquatting projects, and one collection was targeted by three different cybersquatting projects. The remaining collections experienced URI theft from only one cybersquatting project each.
URI theft is relatively uncommon in cybersquatting among NFT collections. This may be due to the high visibility of exact URI duplication, which increases the risk of detection and prompt removal from platforms.

\subsubsection{Image Theft}
We employed \textit{DHash}~\cite{dhash} to evaluate the similarity between images, a method commonly used for detecting near-duplicate or identical images. The similarity between two images is generally measured by the Hamming distance between their \textit{DHash} values. A Hamming distance of less than 10 is typically considered similar, while a distance of less than 5 indicates near-identical images. For our analysis, conservatively, we set the threshold as 5 to identify highly similar images between official and cybersquatting NFT collections. Remarkably, our analysis uncovered 1,483,216 pairs of identical images across 208 official NFT collections and 357 cybersquatting ones, all sharing the same \textit{DHash}. This highlights the widespread use of image duplication as a tactic for deceiving buyers into purchasing counterfeit NFTs.
Moreover, we have identified existing similar images between 130 official NFT collections and 264 cybersquatting ones. Compared to directly using official URIs, attackers prefer to use official images or create similar images to evade detection.

\begin{figure}
    \centering
    \includegraphics[width=0.9\linewidth]{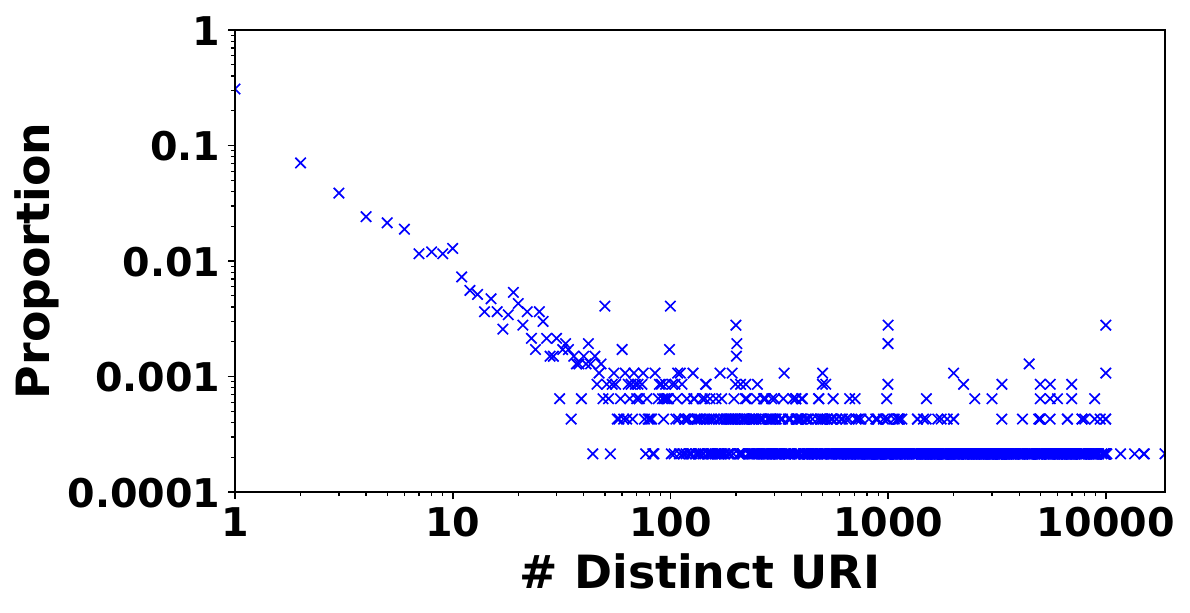}
    \caption{The distribution of cybersquatting NFT collections according to the number of distinct URIs.}
    \label{fig:uri}
\end{figure}
We further analyzed the number of distinct URIs in cybersquatting NFTs. As illustrated in Figure~\ref{fig:uri}, the number of distinct URIs varies significantly. We found that 2,489 cybersquatting NFTs reuse URIs, meaning multiple NFTs are linked to the same URI. For instance, the \texttt{Goblintown Bear Yacht Club} has only 40 distinct URIs, yet it is used by 1,020 NFTs. Our findings revealed that 67\% (52 of 77) of the NFTs involved in URI theft and 66\% (238 of 357) of the NFTs involved in image theft also employ this strategy. This URI reuse strategy reduces production costs, making it easier and cheaper to produce large quantities of counterfeit NFTs. 

\textbf{\textit{Findings:}}
Compared to directly using official URIs, attackers prefer to use official images or create similar images to evade detection, and they also reuse URIs to further lower their costs.

\subsection{Social Media Analysis}
\label{sec:social_media}
In the NFT ecosystem, social media plays an essential role in maintaining project visibility and attracting new investors~\cite{kapoor2022tweetboost}. Legitimate NFT projects often rely heavily on social media to sustain engagement, generate continuous interest, and attract a growing audience of participants.
Among the social media platforms, Twitter is the most commonly used for promoting NFT projects, where both Discord and external promotional websites also serve as important secondary channels.
Prior research~\cite{zhang2024leveraging, luo2022understanding} has demonstrated that Twitter and Discord are particularly effective for NFT promotion due to their real-time interaction capabilities and community-driven dynamics.
In this part, we will characterize the role of social media in cybersquatting NFT collections.

\subsubsection{Twitter}
\begin{figure}
    \centering
    \includegraphics[width=0.9\linewidth]{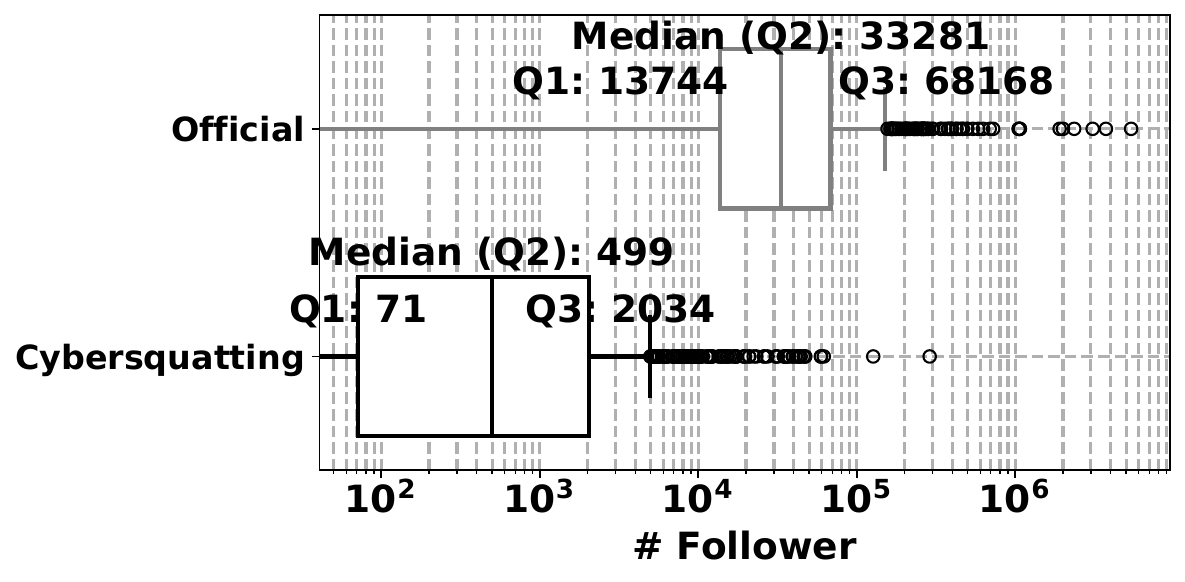}
    \vspace{-0.05in}
    \caption{The comparison of the number of Twitter followers between official and cybersquatting NFT collections.}
    \label{fig:Followers}
\end{figure}

\begin{figure}
    \centering
    \includegraphics[width=0.9\linewidth]{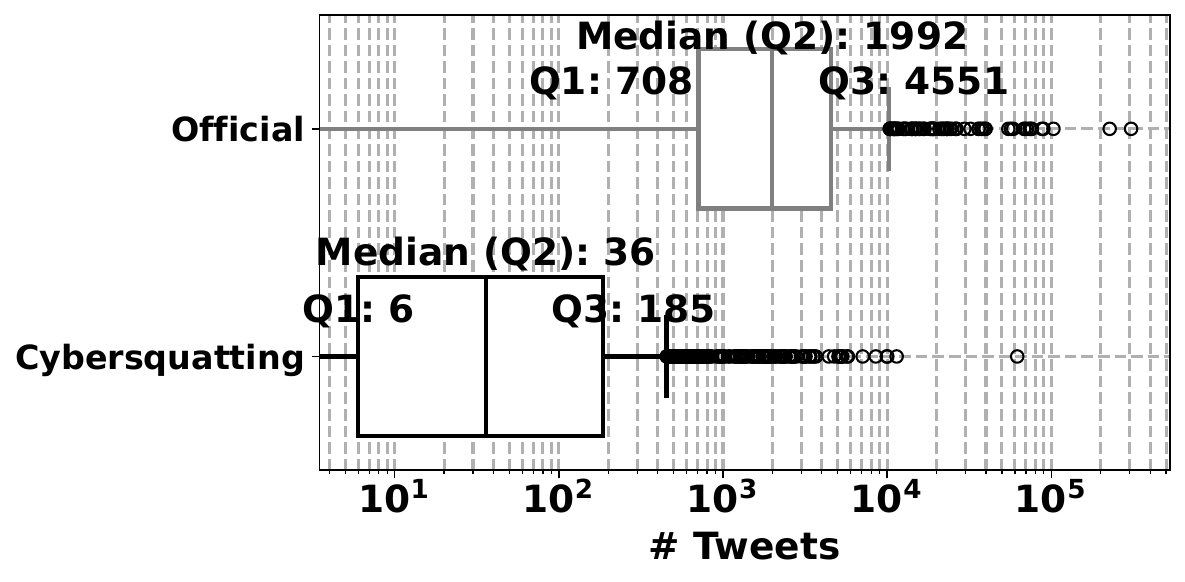}
    \vspace{-0.05in}
    \caption{The comparison of the number of tweets between official and cybersquatting NFT collections.}
    \label{fig:Tweet}
\end{figure}

Twitter has become the dominant platform for NFT promotion, offering real-time updates, community engagement, and direct communication with potential investors. 
From the cybersquatting NFT collections we analyzed, we identified 1,479 distinct Twitter accounts associated with 1,594 collections. Of these, only 966 accounts were still active at the time of writing.
Figure~\ref{fig:Followers} and \ref{fig:Tweet} compare the official and cybersquatting NFT collections in terms of the number of Twitter followers and tweet, respectively.
As we can see, for cybersquatting NFT collections, 75\% have fewer than 2,034 followers, and 25\% have under 499. In contrast, 75\% of official collections have more than 13,744 followers. Similarly, 75\% of cybersquatting collections have fewer than 185 tweets, with 25\% posting fewer than 36. For official collections, 75\% have over 708 tweets.
This highlights the consistently greater vitality of official NFT collections over cybersquatting ones.

We also focused on the distribution of tweets.
We observed that only 205 of the 966 active Twitter accounts associated with cybersquatting NFT collections had posted in the past month, meaning nearly 80\% of them had been inactive for over a month, though their accounts are not suspended.
This stark contrast in social media engagement highlights the superficial nature of cybersquatting NFT collections compared to the official ones, which consistently manage and actively engage with their followers to maintain long-term interest.
The results indicate that cybersquatting NFTs are primarily focused on short-term scams, aiming to deceive quickly and disappear. Unlike legitimate projects that build and nurture their communities, cybersquatting NFT creators show little effort to maintain an ongoing presence or engage with their audience.

\subsubsection{Discord \& External Link}
\label{sec:exteranl_link}

In our analysis, we found only 650 Discord links associated with cybersquatting NFT collections, and of these, only 72 were still active.
The majority of the links had expired, again reflecting the short-term nature of these scams. Additionally, we discovered 2,278 external websites linked to cybersquatting NFT collections.
Some NFTs directly used legitimate project website links in their metadata to evade detection and increase the likelihood of deceiving users, and we excluded these from our analysis. Testing links for accessibility, we found that only 545 were still functional. Beyond accessibility, we scanned the websites using VirusTotal~\cite{VirusTotal}, which flagged 889 sites (39\%) as malicious.
For example, \texttt{azukix.com} was flagged as phishing~\cite{azukix_flag}, originating from the project \texttt{AzukiX}, a mimic of the official NFT collection \texttt{Azuki}. This indicates that many of these websites were being used for phishing or other malicious activities.

\begin{figure}
    \centering
    \begin{minipage}{0.47\linewidth}
        \centering
        \includegraphics[height=4.2cm]{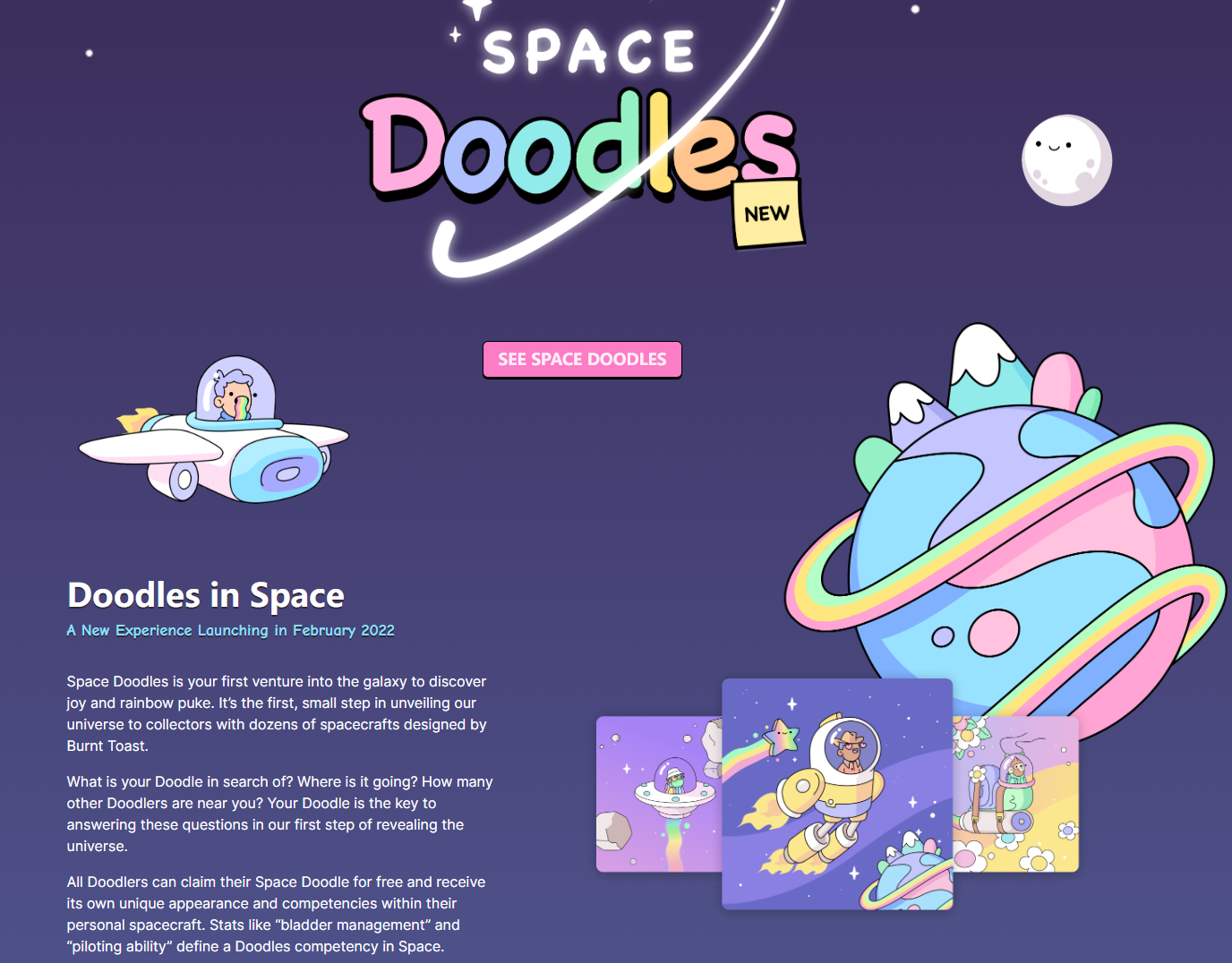}
    \end{minipage}
    \hfill
    \begin{minipage}{0.3\linewidth}
        \centering
        \includegraphics[height=4.2cm]{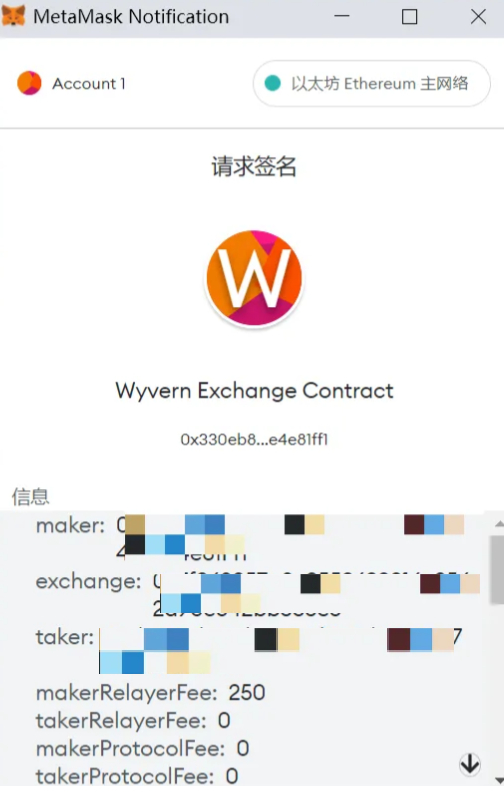}
    \end{minipage}
    \caption{A phishing website and the invoked malicious signature reported by Slowmist.}
    \label{fig:case}
\end{figure}

\noindent \textbf{NFT Phishing.}  
Cybersquatting NFTs involve deceptive naming tactics designed to mislead investors into purchasing counterfeit assets. This technique is often exploited in NFT phishing, where scammers create fake NFTs and embed malicious links in metadata or promotional materials to lure users into visiting phishing websites. While cybersquatting NFTs primarily focuses on fraudulent sales, phishing scams leverage this tactic to compromise users’ wallet credentials, private keys, or seed phrases, resulting in long-term financial losses that extend beyond the initial purchase.
A notable example of NFT phishing is \texttt{https://thedoodles.site}, which was found in nine cybersquatting NFT collections named Space Doodles, impersonating the official Doodles project. The site cloned the official Doodles project website’s design and URL, using subtle visual and URL similarities to deceive users.
Users who visited this phishing site were tricked into signing malicious transactions (see Figure~\ref{fig:case}), allowing scammers to acquire NFTs at minimal cost. Security analysis by Slowmist~\cite{slowmist}, a well-known Web3 security company, confirmed that this site was indeed a phishing site designed to steal user assets\footnote{\href{https://slowmist.medium.com/how-scammers-are-paying-nothing-for-your-nfts-5dd5fe594178}{https://slowmist.medium.com/blog}}. 

To further characterize features of NFT phishing, we randomly sampled 30 websites from the 889 flagged as malicious by VirusTotal. These 30 sites were manually analyzed based on their visual design, domain names, and URL similarities. Given that only 109 are still active, we take advantage of WebArchive~\cite{webarchive} to analyze the inactive websites. Unfortunately, as WebArchive cannot load and execute JavaScript, we were unable to observe wallet malicious signing behaviors on these archived pages.
Our analysis revealed that all 30 selected websites closely mimicked the UI and UX style of legitimate NFT project websites. All these phishing sites used domain squatting techniques to create URLs resembling the official project URLs. This indicates that these sites were specifically designed to deceive users by appearing nearly identical to legitimate NFT projects.

While traditional cybersquatting scams typically focus on obtaining user credentials for later use in offline fraud, NFT phishing allows attackers to directly steal valuable assets through on-chain transactions, which leads to one obvious difference: NFT phishing websites are typically shorter-lived.
Following the taxonomy in Yang~\etal~\cite{yang2024stole}, these NFT phishing sites can be further categorized:
(1) Deceptive Signature, where users are lured into signing seemingly benign messages (\textit{e.g.,} via a fake \texttt{Mint} button) that actually authorize NFT listings at extremely low prices, enabling attackers to acquire assets with little cost (\textit{e.g.,} \texttt{thedoodles.site});
(2) Fraudulent Authorization, where users are tricked into approving NFT transfer permissions to attacker-controlled addresses, often through misleading interface prompts that disguise the true nature of the transaction (\textit{e.g.,} \texttt{pepehedz.xyz});
and (3) Credential Theft, where victims are induced to input their seed phrases or private keys into fake wallet interfaces that mimic legitimate platforms, allowing attackers to take full control of their wallets (\textit{e.g.,} \texttt{xliens.xyz}).

\textbf{\textit{Findings:}}
All data acquired from social media indicates that creators of cybersquatting NFTs show no intention of maintaining long-term operations. Notably, 49.23\% of cybersquatting NFTs lack any of the three social media platforms mentioned. Among those with social media presence, 34.68\% of Twitter accounts are suspended, 88.92\% of Discord links are expired, and 39\% of external websites are flagged as malicious.
Some scammers have used NFTs as a vehicle for distributing phishing links.

\begin{framed}
\noindent\textbf{Answer to RQ2:}
\textit{Compared to legitimate NFTs, cybersquatting NFT collections typically have a lower total supply, shorter active periods, and significantly less social media engagement.
These collections are designed for short-term profit, with scammers showing little intention to maintain long-term operations.
Furthermore, content theft, particularly of images, is a common tactic employed by scammers to deceive users while reducing the cost of creating counterfeit NFTs. This combination of low supply, brief activity, and stolen content enables them to quickly mislead investors before disappearing from the market. }
\end{framed}

\section{RQ3: Actors \& Profits}
\label{sec: actors_profits}
In this section, we first characterize the actors behind all these identified cybersquatting NFT collections, \textit{i.e.,} \textit{scammers} and \textit{victims}. Then, we try to quantitatively depict how much profits are obtained by scammers by focusing on mint fees and creator earnings.

\subsection{Scammer}
\label{sec:tracking_criminal_team}
As illustrated in \S\ref{sec:stakeholders}, scammers are the creators of cybersquatting NFT collections.
Overall, the 8,019 cybersquatting NFT collections are initiated by 6,411 distinct scammers. Notably, 11.44\% (734 out of 6,411) of them were involved in releasing more than one collection.
This finding points to organized criminal groups specializing in producing cybersquatting NFTs.
To identify if scam campaigns exist, we design a three-phase clustering method, \ie, \textit{external link clustering}, \textit{creator clustering}, and \textit{depositor clustering}, to identify scam campaigns.

Specifically, in external links, NFT collections may deploy dedicated websites for promotion. Thus, we heuristically take NFT collections with the same external links as a cluster. 
Note that we do not take external links that appear in official NFT collections into consideration to avoid false positives. By excluding official links, we filter out cases where scammers copy official URLs to impersonate legitimate collections. The remaining projects, with unique external links, are designed by the creators, so collections pointing to the same link are considered part of the same scam team.
In the second phase, we consider the address of the creators. If two collections are created by the same address, we will merge the corresponding two clusters. 
Finally, in the last phase, we take the reuse of deposit addresses of exchange into consideration, which is adopted as a heuristic by a previous work~\cite{victor2020address}.
Specifically, the deposit address is unique to each user and is used to receive funds that are later transferred to the DEX for trading. Addresses transferring funds to the same deposit address are typically linked to the same user. 

The identification of deposit addresses is detailed below.
We first crawled exchange addresses from two platforms: Etherscan~\cite{etherscan}, which provides detailed address labels where exchange addresses are marked as \textit{Exchange}, and Coincarp~\cite{coincarp}, a cryptocurrency data provider. 
Then, we collected transaction data from block 6,000,000 (July 2018) to block 18,050,000 (September 2023) using the Geth client node. Following the method of Victor~\etal~\cite{victor2020address}, we identified deposit addresses that serve as intermediaries between NFT creators and exchanges. Typically, a deposit address exhibits the following three key characteristics: (1) direct transaction relationships - receiving funds from creators and forwarding them to exchanges, (2) minimal fund retention - the amount received from creators approximately equals the amount forwarded to exchanges with minimal losses, and (3) quick forwarding - completing the forwarding within a short time period.
Based on these characteristics, we identify deposit addresses by requiring the amount difference between received and forwarded funds to be less than 0.01 Ether, and the forwarding to be complete within 10,000 blocks. To further explore relationships between creators, we expand the address dataset by parsing the \textit{from} and \textit{to} fields in transactions involving the creator, increasing the address set for clustering. We then perform deposit-based clustering by grouping addresses that interact with the same deposit address, indicating a connection between them.

In total, 794 clusters were identified, involving 3,811 NFT collections and 2,203 creators.
Notably, approximately 90\% of the clusters contain fewer than 8 addresses, while only 8 clusters have 50 or more addresses.
Interestingly, we identified two distinct types of clusters: 317 clusters where all NFTs point to the same external link, and 382 clusters without any external links, primarily centered around a single creator address.
Table~\ref{tab:top5Cluster_link} shows the five largest clusters of the first type. We noticed that they all point to a malicious external website that is flagged by VirusTotal~\cite{VirusTotal}.
Further analysis on all these 317 clusters reveals that these clusters typically target a single official NFT collection and often involve multiple creators. Many of the external links used are no longer accessible, and a significant portion has been flagged as phishing by VirusTotal.
In contrast, the second type of cluster focuses on producing cybersquatting NFTs targeting various official projects. As shown in Table~\ref{tab:top5cluster_creator}, they are all created and deployed by a single creator but target several NFT collections, thereby increasing the potential number of victims.

As for the largest cluster, it includes 206 cybersquatting NFT collections targeting 95 official NFT collections, with 53 external links, 18 of which have been flagged as malicious by VirusTotal.
This cluster is a unique case, clearly demonstrating how scammers can adopt a mixed strategy, combining tactics from both types of clusters. Specifically, this cluster targets multiple official projects while actively leveraging phishing links to deceive users. This suggests that in some cases, scammers may deliberately employ both phishing-based and multi-project targeting approaches within a single campaign, though this is not the general rule across all clusters.

\textbf{\textit{Findings:}} 
According to our three-phase clustering method, we have identified 794 clusters, which may correspond to scam campaigns. Two types of scam campaigns widely exist, \textit{i.e.,} clusters centered around a single external link and clusters centered around a single creator.
The first type often involves a malicious link, aiming to use NFTs as a means to spread phishing websites. The second type focuses on creating as many cybersquatting NFTs as possible to increase the number of potential victims and raise the success rate of the scam.

\begin{table*}[t]
\centering
\caption{Top-5 scam campaigns deployed by the same creator address.}
\label{tab:top5cluster_creator}
\begin{tabular}{@{}cccc@{}}
\toprule
\textbf{Creator Address}                   & \begin{tabular}[c]{@{}c@{}}\textbf{\#Cybersquatting}\\\textbf{NFT Collections}\end{tabular} & \begin{tabular}[c]{@{}c@{}}\textbf{\#Targeted}\\\textbf{NFT Collections}\end{tabular} & \begin{tabular}[c]{@{}c@{}}\textbf{Targeted}\\\textbf{NFT Collections}\end{tabular}                                                                         \\ \midrule
0xd973564a85ee827e7f983c9eaacadd6fa74b9da1 & 67              & 62                    & \begin{tabular}[c]{@{}c@{}}Wolf Game, Cool Pets, Azuki...\end{tabular}                     \\
0x90ea805e049c06bdf233c8d3754707f8b2054673 & 57              & 53                    & \begin{tabular}[c]{@{}c@{}}AzMURI, mfer, Milady...\end{tabular}                            \\
0xdf57686394c637e38c05e595df31c58d25d8e50c & 56              & 51                    & \begin{tabular}[c]{@{}c@{}}Mutant Ape Yacht Club, Wolf\\ Game, World Of Women...\end{tabular} \\
0x945dc4a0e40b4fb183389a0f3e6ebd100819e276 & 44              & 38                    & \begin{tabular}[c]{@{}c@{}}Wolf Game, 10KTF, CloneX...\end{tabular}                        \\
0xf8238a3dd9a67b8419412ede613a06d73ffc2d93 & 45              & 35                    & \begin{tabular}[c]{@{}c@{}}Milady, Wolf Game, Nouns...\end{tabular}                        \\ \bottomrule
\end{tabular}%
\end{table*}

\begin{table*}[t]
\centering
\caption{Top-5 scam campaigns that mimic the same official NFT collection.}
\label{tab:top5Cluster_link}
\begin{tabular}{@{}cccccc@{}}
\toprule
\textbf{External Link}  & \begin{tabular}[c]{@{}c@{}}\textbf{\#Cybersquatting}\\\textbf{NFT Collections}\end{tabular} & \multicolumn{1}{l}{\textbf{\#Creator}} & \begin{tabular}[c]{@{}c@{}}\textbf{\#Targeted}\\\textbf{NFT Collections}\end{tabular} & \begin{tabular}[c]{@{}c@{}}\textbf{Targeted}\\\textbf{NFT Collections}\end{tabular} & \textbf{Official Website}                                                 \\ \midrule
goblintown.link         & 56               & 56                                      & 1                      & goblintown          & goblintown.wtf                                                            \\
killabears.in           & 50               & 50                                      & 1                      & KILLABEARS          & killabears.com                                                            \\
shinsekai.link          & 50               & 50                                      & 1                      & Shinsekai           & shinsekai.io                                                              \\
shinsekaidrifters.link  & 50               & 50                                      & 1                      & Shinsekai           & shinsekai.io                                                              \\
murakamiflowersmeta.xyz & 49               & 49                                      & 1                      & Murakami.Flowers    & \begin{tabular}[c]{@{}c@{}}murakamiflowers.\\ kaikaikiki.com\end{tabular} \\ \bottomrule
\end{tabular}%
\end{table*}

\subsection{Victim}
\label{sec:victim}
\begin{figure}[t]
    \centering
    \includegraphics[width=0.9\columnwidth]{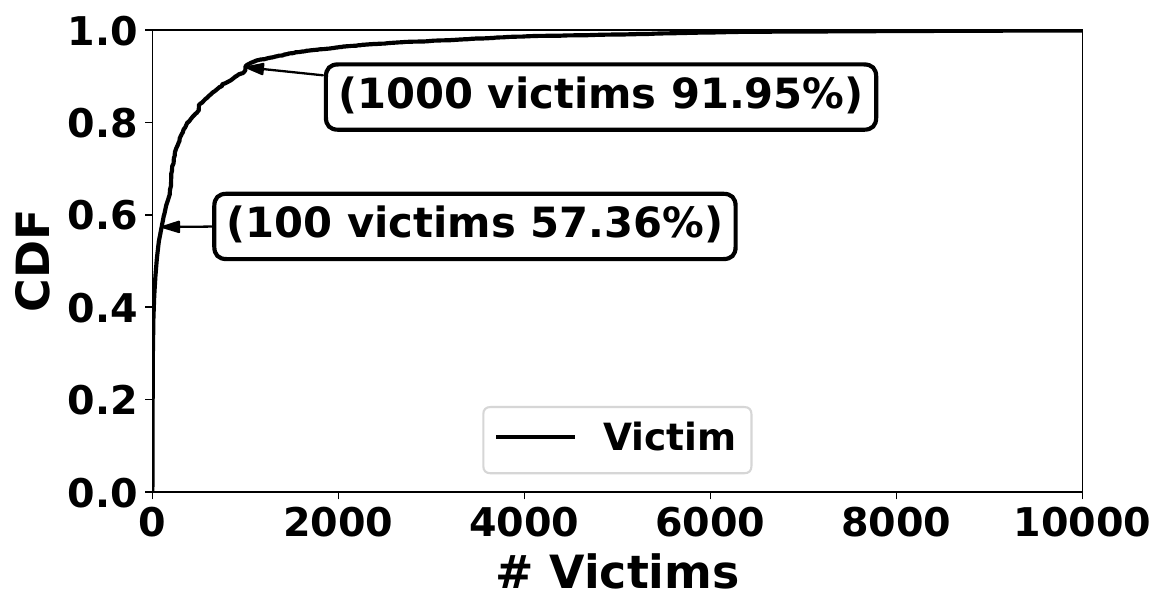}
    \caption{The CDF of cybersquatting NFT collections according to the number of victims.}
    \label{fig:CDF_victim}
\end{figure}

\begin{figure}[t]
    \centering
    \includegraphics[width=0.9\columnwidth]{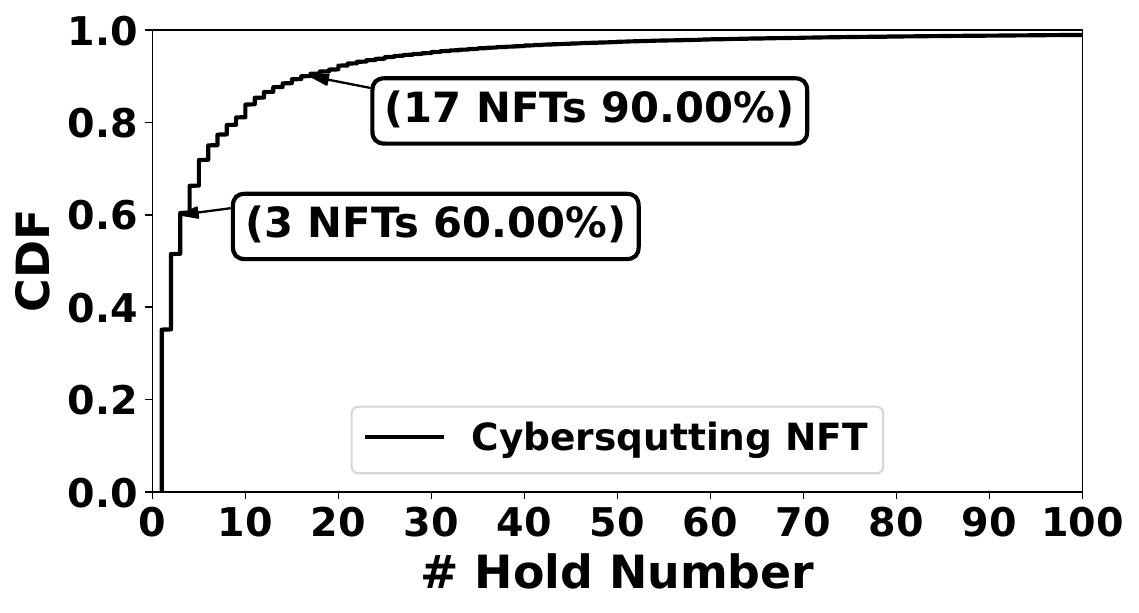}
    \caption{The CDF of cybersquatting NFT collections according to the number of held NFTs.}
    \label{fig:CDF_hold}
\end{figure}

As illustrated in \S\ref{sec:stakeholders}, victims in cybersquatting NFT schemes include both minters and buyers, as they directly incur financial losses. Out of 782,082 involved addresses in both NFT transfers and trades, we identified 670,817 unique victim addresses.

We first illustrate the distribution of victims among cybersquatting NFT collections, as shown in Figure~\ref{fig:CDF_victim}. As we can see, 57.36\% of cybersquatting NFT collections have no more than 100 victims, and 91.95\% of them have fewer than 1,000 victims. Notably, only six collections have over 10,000 victims, which have been listed in Table~\ref{tab:top_victim}.
The most affected collection is \texttt{Mindblowon}, which is mimicked by \texttt{MindBlowon}~\footnote{\href{https://etherscan.io/token/0xc1906d4eebd43039d21970de0724b411c713c563}{https://etherscan.io/token/0xc1906d...}}, which has been flagged as spam by Etherscan already~\cite{etherscan}. The fourth collection in the table, \texttt{Hobgoblintown}, has not been labeled as either spam or phishing, though it is still a cybersquatting NFT, using combination squatting to impersonate the official NFT collection \texttt{goblintown}. The other four collections have also been marked as malicious, either as spam or phishing. These cybersquatting NFTs impersonate official collections, often conducting phishing attacks by airdropping large quantities of NFTs with embedded phishing links to users.
According to our statistics, we found that the top 10\% of collections account for more than 70\% of the total number of victims. This demonstrates a serious issue, where those top collections attract the overwhelming majority of victims.

We further investigated the number of cybersquatting NFTs held by individual victims, as shown in Figure~\ref{fig:CDF_hold}. We found that 60\% of victims hold only three cybersquatting NFTs, while over 90\% hold no more than 17 cybersquatting NFTs.
The data shows that most victims do not hold many cybersquatting NFTs, which suggests they may be deceived occasionally due to limited experience in trading cryptoassets.
We also checked the transaction status and ETH balance of their accounts. We found that 90\% of victim accounts had a balance of no more than 0.2 ETH, and more than half had fewer than 100 transactions, further reinforcing their lack of trading experience.

\textbf{\textit{Findings:}} The analysis of victims reveals that most cybersquatting NFT collections have a relatively small number of victims, while a small percentage of collections account for the majority of victims.  Victims are novices with limited investment in NFTs, making them vulnerable to cybersquatting scams.

\begin{table*}[t]
\centering
\caption{Top-6 cybersquatting NFT collections ranked by the number of victims.}
\label{tab:top_victim}
\begin{tabular}{@{}lcccc@{}}
\toprule
\textbf{NFT Collection Name} & \textbf{Contract Address}                  & \begin{tabular}[c]{@{}c@{}}\textbf{Targeted NFT}\\\textbf{Collection Name}\end{tabular} & \textbf{\#Victims} & \textbf{Label} \\ \midrule
MindBlowon    & 0xc1906d4eebd43039d21970de0724b411c713c563 & Mindblowon      & 21,877               & Spam             \\
Momoco Azuki  & 0x5838f7c3d22da78d8f473130ed80ed07dd1df2eb & Azuki           & 17,593               & Spam          \\
FlowerFam     & 0xbfb3e3e67cfebe05c53967c9f0ecda09e06658b3 & Flower Fam      & 13,587               & Spam          \\
Hobgoblintown & 0xe55d6a095adf0f6eae251169d9ab6e406d0695cd & goblintown      & 11,702               & -         \\
MindBlowon    & 0x68d00e0da009e4322dc1ab33100ba4ccc89b9fcc & Mindblowon      & 11,437               & Spam             \\
MutantRatsNFT & 0xbcc3d1b4ddf3b9d38e7591c917d4db9501ecc3d6 & MutantCats      & 10,480               & Phishing         \\ \bottomrule
\end{tabular}%
\end{table*}

\subsection{Mint Fees \& Creator Earnings}
As illustrated in \S\ref{sec:stakeholders}, both mint fees and creator earnings can be taken as profits. In this section, we will investigate those associated with cybersquatting NFT collections.
We calculated the total mint fees by identifying mint events(see \S\ref{sec:transfer_event}) for each collection, summing the ETH involved in these transfers. Creator earnings were determined by multiplying the royalty percentage from our NFT metadata by the sale price on the secondary market and then summing the results for each collection.

An NFT collection is defined as having profit if it has generated earnings from either mint fees, creator earnings, or both. There are only 2,255 cybersquatting NFT projects profitable. 
We observed that they collectively generated profits amounting to \texttt{21.6K ETH}, approximately \texttt{\$59.26 million}. This profit was derived either from mint fees, creator earnings, or both. Specifically, 461 projects profited solely from creator earnings, while 935 projects gained exclusively from mint fees. A smaller subset, 859 projects, benefited from both income streams. Notably, a substantial proportion (71.87\%) of these cybersquatting NFTs did not realize any profit. And the discrepancy in earnings between the most and least profitable NFT projects reaches a staggering factor of $10^{18}$, and the top 10\% of profitable NFTs captured 84.64\% of the total profit.

As for mint fees, we found that 1,794 (1,794/8,019, 22.37\%) cybersquatting NFT collections have mint fees.
These collections have made cumulative profits up to \$53.35M (19.35K ETH) from minting fees. However, over half (967/1,794) of the collections made a profit of no more than 1 ETH, and the total profit from these collections is only 3.7\% of the total.
Astonishingly, 20\% of projects capture over 90\% of mint fee revenue.
As shown in Table~\ref{tab:Top5mintfee}, the most profitable project made up 487.61 ETH via mint fees. 
We further studied these collections in the table. We found that the names of these cybersquatting NFTs were created through the combination squatting strategy, and their linked pictures are extremely similar to the style of the official NFT collection.
At the same time, the creators also set up social media accounts and external promotional websites. It can be seen that the scammers of these five projects have invested a lot of cost and effort, and therefore have obtained higher profits.
However, the Twitter accounts of these projects have stopped updating, the corresponding Discord links have expired, and the external promotional websites are inaccessible.
This shows that the purpose of creating these cybersquatting NFT collections is to make profits in the short term, which is consistent with the previous observations.

As for creator earnings, we observed that 1,320 (16.46\%) cybersquatting NFT collections generate profits through this channel. These projects have accumulated total profits of up to \$5.9M (2.25K ETH) from creator earnings.
The top 5\% of projects account for over 99\% of creator earnings. Notably, there is a substantial disparity in profitability among projects, with \texttt{Grandpa Ape Country Club} being the most lucrative, earning up to 176.95 ETH. This highlights the significant variation in profit margins among different projects in the cybersquatting NFT landscape. As shown in Table~\ref{tab:top5CreatorEarn}, the most profitable project achieved 176.95 ETH creator earnings.

\begin{table*}[t]
\centering
\caption{Top-5 cybersquatting NFT collections ranked by obtained mint fees.}
\label{tab:Top5mintfee}
\begin{tabular}{cccc}
\toprule
\begin{tabular}[c]{@{}c@{}}\textbf{Cybersquatting NFT}\\\textbf{Collection Name}\end{tabular} & \begin{tabular}[c]{@{}c@{}}\textbf{Targeted NFT}\\\textbf{Collection Name}\end{tabular} & \textbf{Contract Address}                  & \textbf{\#ETH (USD)} \\ \midrule
Timeless Ape Club             & Timeless        & 0xd95c08db87628febfc551c1ff1cfc9fe1269ef9c & 487.61 (1.36M)        \\
Doodles Flipped               & Doodles         & 0x98b486f4fd2a1526eb6fd09f200735d4a9fcadfa & 372.72 (1.16M)        \\
The Plague                    & The Plague      & 0x8c3fb10693b228e8b976ff33ce88f97ce2ea9563 & 359.98 (1.07M)       \\
Metaverse Cool Cats           & Cool Cats       & 0xfa8e23f3d18d8d672cd4465b0bd7af81d55cf2b6 & 349.19 (1.3M)        \\
Azuki Mfer                    & Azuki           & 0xee467844905022d2a6cc1da7a0b555608faae751 & 306.63 (0.82M)      \\ \bottomrule
\end{tabular}
\end{table*}

\begin{table*}[t]
\centering
\caption{Top-5 cybersquatting NFT collections ranked by obtained creator earnings.}
\label{tab:top5CreatorEarn}
\begin{tabular}{cccc}
\toprule
\begin{tabular}[c]{@{}c@{}}\textbf{Cybersquatting NFT}\\\textbf{Collection Name}\end{tabular} & \begin{tabular}[c]{@{}c@{}}\textbf{Targeted NFT}\\\textbf{Collection Name}\end{tabular}          & \textbf{Contract Address}                  & \textbf{\#ETH (USD)} \\ \midrule
Grandpa Ape Country Club      & Grandpa Ape Country Club & 0x656b9e24de2e41a94a7dbbaeb3937777cf34e448 & 176.95 (518K)        \\
mfer chicks                   & mfer                     & 0xda858c5183e9024c0d5301ee85ae1e41dbe0f880 & 103.10 (277K)        \\
ape mfer                      & mfer                     & 0xd629fe374143b60ff4b0decb81673ee85a977d17 & 79.87 (218K)         \\
xmfers                        & mfer                     & 0xb156adf8523fdc6152affdba076a2143fd7e3c69 & 68.81 (182K)         \\ 
Never Fear Truth        & Never Fear Truth   & 0x399bd363426dfdd7f84fe8b917e1a3525b039309 & 63.60 (173K)       \\ \bottomrule
\end{tabular}%
\end{table*}

\textbf{\textit{Findings:}} Our analysis reveals that a small number of top cybersquatting NFT projects captured the majority of profits. 10\% of profitable NFTs captured over 80\% of the total profit.

\begin{framed}
\noindent\textbf{Answer to RQ3:}
\textit{
There is indeed evidence of organized scam campaigns, and we have identified 794 associated groups impacting over 670K victims, who often lack blockchain expertise.  
We further quantified that cybersquatting NFT collections generate profits primarily through two channels: \textit{mint fees} and \textit{creator earnings}, where the total profit obtained by scammers can exceed \$59.26 million.
}
\end{framed}

\section{Discussion}

\noindent \textbf{\textit{Mitigation Strategies.}} 
From the perspective of NFT marketplaces, combining our findings, we have proposed four possible concrete mitigation strategies as follows. 
First, NFT marketplaces should enforce stricter listing policies, such as mandatory creator verification, cryptographic proof of ownership, and deposit fees to prevent the existence of mass scams.
Second, NFT marketplaces should integrate and improve a user reporting system by incorporating crowdsourced intelligence and prioritizing reports based on the reputation scores of users. In this system, users are able to issue real-time risk alerts, highlight suspicious trading patterns, and mark abnormal social media activities.
Third, based on the user reporting system, NFT marketplaces should build a cross-marketplace intelligence database to prevent detected scams from being relaunched elsewhere. All historical scams should be formalized and inserted into this database to facilitate subsequent queries.
Last but not least, NFT marketplaces can take advantage of our real-time detection pipeline, which can cross-reference new NFT collections with a known database of names and on-chain data of historical scams, instantly identifying possible cybersquatting attempts. 

\noindent \textbf{\textit{Lessons for Stakeholders.}}
Except for NFT marketplaces, two other roles exist in the whole ecosystem, i.e., investors and creators (see Figure~\ref{fig:stakeholders}).
Specifically, \textbf{NFT investors} should become aware of common cybersquatting patterns (see \S\ref{sec:name_tactic}) and monitor key indicators such as total supply, social media activity, and transaction volume. Since over 80\% of cybersquatting projects remain active for less than a day and involve fewer than 100 transactions (see \S\ref{sec:basic_feature}), vigilance is crucial when purchasing NFTs. 
Moreover, \textbf{NFT creators} should carefully choose names that avoid resemblance to established projects and monitor the market for imitations (see \S\ref{sec:name_tactic}). They should also remain vigilant against content theft, as there are over 1.4 million cases of duplicate content (see \S\ref{sec:content_theft}), and report infringements to protect their creations.

\noindent \textbf{\textit{Limitations.}}
Firstly, our research is confined to examining cybersquatting NFTs associated exclusively with the top 1,000 official NFTs on the Ethereum blockchain. Although it is plausible that there may be cybersquatting tokens targeting official NFTs outside our study, our observations indicate a higher likelihood of attackers focusing on prominent tokens with a substantial market capitalization rank.
Secondly, we employ domain squatting tools to generate names for NFT squatting. It's important to note that this approach may not encompass all squatting patterns within the NFT domain, and we leave this area for future investigation. It's worth acknowledging that false positives cannot be eliminated. Additionally, since these names are randomly generated, they may not encompass all instances of cybersquatting NFTs.
Lastly, NFTs have gained widespread traction on various blockchain platforms beyond Ethereum, such as BNB Smart Chain, Solana, and others. We maintain that our proposed methodology can be adapted and applied to cybersquatting NFTs on these alternative platforms.

\section{RELATED WORK}
\noindent \textbf{Domain Squatting.} There is a lot of research on domain squatting that analyzes it from the perspective of the technology it uses. Spaulding \etal ~\cite{spaulding2016landscape} systematically studied typosquatting (squatting via typographical errors).  Nikiforakis \etal ~\cite{nikiforakis2013bitsquatting,nikiforakis2014soundsquatting} studied bitsquatting (squatting via bit flips) and sound squatting (abuse of the pronunciation similarity of different words).  
Gabrilovich \etal ~\cite{gabrilovich2002homograph} and Holgers \etal ~\cite{holgers2006cutting} studied homograph-based squatting(squatting via the use of characters from different sets). 
Kintis \etal ~\cite{kintis2017hiding} studied combosquatting (combination name with other keywords). Spaulding \etal ~\cite{spaulding2016landscape} provides a summary of the various existing domain squatting techniques. The use of domain technology has already been found in other fields. Hu \etal~\cite{hu2020mobile} studied the domain squatting phenomenon in mobile apps. Xia \etal ~\cite{xia2022challenges} studied a similar phenomenon in the ENS ecosystem. We take the first step to characterize squatting in the NFT ecosystem.

\noindent \textbf{Non-fungible Token (NFT).} NFT has only really exploded in the last year, and studies of its own specificity and ecology have been emerging. Park \etal ~\cite{park2022evolution} studied the complexity and novelty of NFT use cases. 
Wang \etal ~\cite{wang2021non} systematically studied the Ethereum NFT ecosystems and highlighted a series of open challenges in NFT ecosystems. 
Huang \etal ~\cite{huang2024unveiling} studied the NFT market behaviors.
Das \etal ~\cite{das2022understanding} systematically analyzed the various security issues in the NFT ecosystem and used the Levenshtein Distance to identify potential similar name NFTs.  But focuses only on names with more than 7 characters, which results in a significant number of false negatives and false positives. Kapoor et al.~\cite{kapoor2022tweetboost} investigated the impact of the important social media platform, Twitter, on the prices of NFTs.
Pinto-Guti{\'e}rrez \etal ~\cite{pinto2022nft} studied the hype against NFT. Ma \etal~\cite{ma2023sok} from the perspective of NFT incidents, investigated the security issues present in the NFT ecosystem.

\noindent \textbf{Cryptocurrency Scams.}
Since the ongoing boom of cryptocurrencies and the increasing market capitalization, all kinds of scams have emerged. Some studies have characterized some types of cryptocurrency scams, including Ponzi schemes~\cite{chen2018detecting,chen2019exploiting,fan2020expose,shen2021mining,mukherjee2021cryptocurrency,bartoletti2018data}, phishing~\cite{andryukhin2019phishing,chen2020phishing,yuan2020detecting,zhang2021blockchain,wu2020phishers}, and scam Initial Coin Offerings (ICOs) \cite{liebau2019crypto,urbaczewski2023predicting,cong2023dark}. 
Gao \etal~\cite{gao2020tracking} studied the counterfeit ERC-20 tokens. Xia \etal ~\cite{xia2020characterizing} characterized cryptocurrency exchange scams. 
Sharma \etal~\cite{sharma2023understanding} characterized the NFT rug pull. 
Huang \etal~\cite{huang2023miracle} devised an effective machine learning-based approach that leverages both on-chain transaction data and off-chain metadata to identify NFT rug pulls by detecting suspicious project behaviors.

\section{CONCLUSION}
This paper has presented the first in-depth measurement study of cybersquatting NFT on the Ethereum blockchain, uncovering 8,019 cybersquatting collections targeting 654 popular projects. We have also uncovered seven distinct squatting tactics, with combination squatting being the most favored by scammers. Our findings expose well-organized criminal groups affecting over 670K victims and demonstrate significant economic impacts with scammers profiting over \$59.26 million. This highlights the urgent need for enhanced detection and prevention strategies within the community to mitigate the risks of NFT cybersquatting. Future efforts should focus on developing effective countermeasures and educating users on the dangers of such scams.

\bibliographystyle{IEEEtran}  
\bibliography{refs} 

\appendix
\section*{Data Availability}
The dataset used in this study has been made publicly available in an anonymized repository to ensure transparency and reproducibility of the research. Access to the dataset can be obtained via the following repository: 
\href{https://github.com/security-pride/Cybersquatting-NFT}{https://github.com/security-pride/Cybersquatting-NFT}.

\section*{Ethical Considerations}
All the data presented in this paper was obtained from the Ethereum blockchain and other publicly available sources. 
We make no attempt to access non-public data.
We exclusively extracted relevant data pertaining to NFTs and the nature of this study. The purpose of sharing this public data is to enable others to utilize the analyzed information and inspire further research on NFTs. 
While our published data may contain instances of malicious activity, which can be studied by potential attackers to conduct further attacks, we believe that studying these existing attacks can provide valuable insights and recommendations to the community.
This can help inspire future research on NFT security measures and prevent similar attacks from occurring.
We have undertaken responsible disclosure to the relevant parties.
We have reported the identified 8019 cybersquatting collections to OpenSea and have received a positive response confirming that they are currently under internal review. By the time of response, 2,572 NFT collections had been removed from the OpenSea platform.
Considering the benefits of identifying these security and privacy concerns, we do not believe our paper raises any ethical concerns.

\end{document}